\documentclass[prl,superscriptaddress,,twocolumn]{revtex4-2}
\usepackage{amsthm,amsfonts}
\pdfoutput=1
\usepackage[english]{babel}
\usepackage{graphicx} 
\usepackage{xcolor}
\usepackage{tikz} 
\usetikzlibrary{shapes,arrows,positioning}
\usepackage{xcolor}
\usepackage{cancel}
\usepackage{braket}
\usepackage{calc}
\usepackage{bbold}
\usepackage{mathrsfs}
\usepackage[colorlinks,citecolor=blue]{hyperref}
\usepackage{url}
\usepackage{color}
\usepackage{arydshln}
\usepackage{float}
\usepackage{framed}
\usepackage{bbold}

\usepackage{amsmath}
\usepackage{soul}
\usepackage[titles]{tocloft}

\usepackage{dcolumn}
\usepackage{bm}
\usepackage[caption=false]{subfig} 

\newcommand{\bpm}{\begin{pmatrix}}
\newcommand{\epm}{\end{pmatrix}}

\definecolor{shadecolor}{gray}{.9}
\def \bes {\begin{subequations}}
\def \eds {\end{subequations}}
\def \beqn {\begin{equation*}}
\def \edqn {\end{equation*}}

\def \dag {\dagger}

\def \up {\uparrow}
\def \down {\downarrow}

\def \sm {\sigma}

\def \epsilon {\varepsilon}

\def \calh {{\cal{H}}}
\def \call {{\cal{L}}}

\def \calp {{\cal{P}}}
\def \calq {{\cal{Q}}}

\def \caln {{\cal{N}}}

\def \calr {{\cal{R}}}

\definecolor{shadecolor}{gray}{.9}
\def \beq {\begin{equation}}
\def \edq {\end{equation}}
\def \bes {\begin{subequations}}
\def \eds {\end{subequations}}
\def \beqn {\begin{equation*}}
\def \edqn {\end{equation*}}

\def \dag {\dagger}

\def \up {\uparrow}
\def \down {\downarrow}

\def \sm {\sigma}

\def \epsilon {\varepsilon}

\def \calh {{\cal{H}}}
\def \call {{\cal{L}}}

\def \calp {{\cal{P}}}
\def \calq {{\cal{Q}}}

\def \caln {{\cal{N}}}

\def \calr {{\cal{R}}}

\def \veps {\varepsilon}

\begin{document}

\title{An optimal superconducting hybrid machine}

\author{Rosa L\'{o}pez}
\email{rosa.lopez-gonzalo@uib.es}
\affiliation{Instituto de F\'{i}sica Interdisciplinar y Sistemas Complejos IFISC (CSIC-UIB), E-07122 Palma de Mallorca, Spain}
\author{Jong Soo Lim}
\email{lim.jongsoo@gmail.com}
\affiliation{Department of molecular design, Arontier Co., 15F Sewon Bldg., Gangnam-daero 241, Seocho-gu, Seoul, Republic of Korea}
\author{Kun Woo Kim}
\email{kunx@cau.ac.kr}
\affiliation{Department of Physics, Chung-Ang University, 06974 Seoul, Republic of Korea}

\begin{abstract}
Optimal engine performances are accomplished by quantum effects. Here we explore two routes towards ideal engines, namely (1) quantum systems that operate as hybrid machines being able to perform more than one useful task  and (2) the suppression of fluctuations in doing such tasks. For classical devices, the absence of fluctuations is conditioned by a high entropy production as dictate the thermodynamic uncertainty relations. Here we generalize such relations for multiterminal conductors that operate as hybrid thermal machines. These relations are overcome in quantum conductors as we demonstrate for a double quantum dot contacted to normal metals and a reservoir being a generator of entangled Cooper pairs. 
\end{abstract}
\maketitle

\emph{Introduction---} The second law of thermodynamics dictates that Carnot efficiency is the maximum efficiency for a thermal machine that delivers zero power and works reversibly \cite{Carnot}. From a practical point of view, however, under a nonequilibrium situation a thermal engine generates some power and a finite amount of entropy. In such a scenario,  a new kind of nonequilibrium principle establishes a connection between power fluctuations and entropy production, which is called the thermodynamic uncertainty relation (TUR) \cite{Barato2015,Todd2016,Pietz2016,Pietzonka16b,Pietz2018,Gingrich17,Pietz2018,
Timp2019,Hase2019,Pietzonka17,Horowitz17,Shiraishi17,Proesmans17,Macieszczak18}.  
The relation  expresses a trade-off between power fluctuations and entropy production. The entropy production imposes a bound for the power fluctuations, suggesting that there may be fundamental limitations on the precision of thermal machines. However, this is only a bound for classical systems since quantum systems are able to circunvent the TUR. The relevance of the TUR has been extended to be related with other important concepts of nonequilibrium thermodynamics, including fluctuation theorems \cite{Timp2019,Hase2019} and information theory \cite{Bija2018,Mars2018,Potts19}.

\begin{figure}[t]
\centering
\includegraphics[width=1\linewidth]{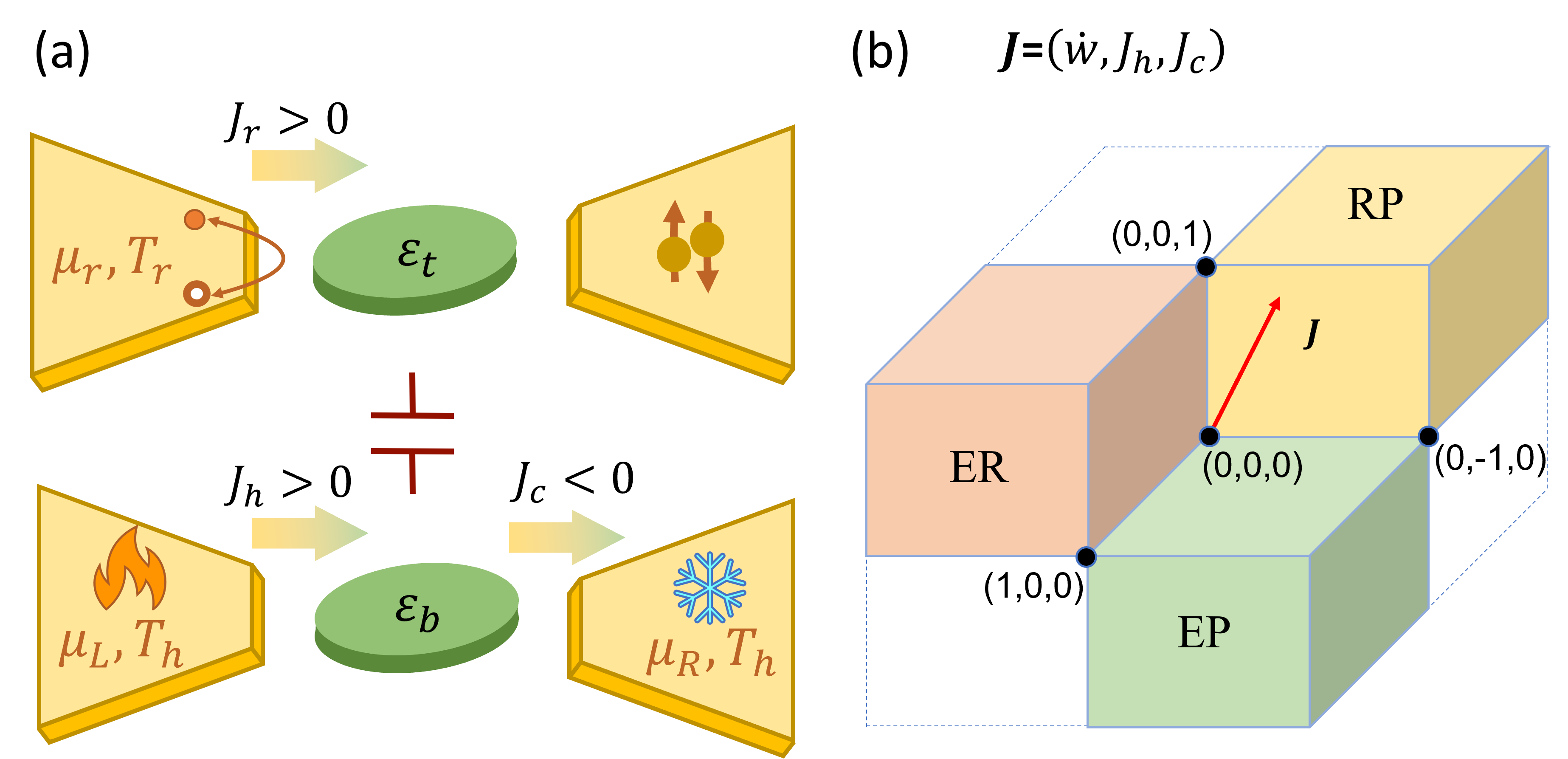}
\caption{(a) Scheme of the double quantum dot setup as a hybrid thermal machine. Two quantum dots are capacitively coupled, and coupled to three normal leads with $T_{r,h,c}$ and one superconducting lead. The heat current from each lead to the system is denoted by $J_{r,h,c}$ and its sign criteria is indicated. 
(b)Three working modes of this device (RP, ER, EP) are indicated in the domain of $\bold J = (\dot w,J_h,J_c)$, where $\dot  w = J_r+J_h+J_c$ means the production rate of work. For instance, when a system is in RP mode (refrigerating and heat pumping), $\dot w<0, J_h<0, J_c>0$. 
The EP mode is generating work and pumping heat,  $\dot w>0, J_h<0, J_c<0$. The heat current $J_{h,c}$ is positive when it flows to the system.
}
\label{fig:1}
\end{figure}

Standard thermal machines consists of a working substance connected to two heat baths, a cold reservoir and a hot one. Then, depending on the performed task they can work as heat engines (E), refrigerators (R) or pumping machines (P). Recently a novel characterization for systems connected to 
more than two terminals, i.e., multiterminal devices, has been formulated. These are hybrid thermal machines and have the advantage of performing more than one useful task simultaneously. For these devices a generalized efficiency has been enunciated  \cite{Gonzalo2020}. Our goal is to extend the TUR for these kind of multiterminal thermal machines and relate its efficiency (for the different working modes listed below)  with its current-current correlations. As previously stated, we illustrate how these multidimensional TUR (MTUR) can be  circumvented when the working substance is a quantum system.  For such case we show that MTUR are violated and its cause is attributed to quantum coherence, to the non-local character of quantum states in entangled states or the breakdown of LDB by quantum effects~\cite{SFR}. Therefore, exploiting  quantumness can be advantageous in designing a precise quantum engine with low dissipation \cite{Ptasz2018,Agar2018,Brand2018,Liu2019} as the generalized MTUR can be overcome. Definitely, the strong bound imposed by the MTUR needs a revision for quantum systems where a careful analysis must be done in view of  coherence,  interactions, or non-Markovian dynamics among others \cite{Horodecki2013,Sega2018,Brandner2018, Ptaszy2018,Sary2019,Buff2020,Bene2020}. 


We propose to deal with a minimal setup consisting in two nanoscale conductors which are capacitively coupled, each of them connected to two baths (see Fig.~\ref{fig:1}). The setup is able to perform useful tasks, for instance, conducting an electrical current, cooling a cold reservoir, or pumping heat when it is electrically or thermal biased (or both). In this sense we are dealing with a hybrid thermal machine \cite{Gonzalo2020}. Our study focus on the bounds obtained from the thermodynamic uncertainty relations that are generalized to treat multiterminal conductors. Departures from these bounds are illustrated with our device in which the presence of coherent states and the breakdown of local detailed balance (LDB) take place. We consider one of the contacts being a superconductor, i.e., a reservoir of coherent Cooper pairs. Besides, each conductor consists of a quantum dot, which is chosen due to its ability as energy filters, a characteristic that potentially enhances its thermoelectrical behavior.


\emph{Hybrid thermal Machines: Efficiency and MTUR---} 
Frequently, the efficiency of a thermal engine is measured by the ratio of the output to the input.  
When more than two terminals are attached to an engine, however, the physical scenario becomes more involved.  In general,  multiterminal devices  may exhibit ill-defined efficiencies when they are characterized in a standard manner~\cite{Hajiloo20}. The main cause is that the engine can produce multiple outputs simultaneously.
This occurs, for example,  when  work is extracted and at the same time heat is pumped  towards the hot reservoir \cite{Gonzalo2020} in three terminal devices, the efficiency $\eta_{EP}$ is for a simultaneous engine-pumping mode.  Other operating modes correspond to refrigerator-pumping (RP) and engine-refrigerator (ER) regimes.
In the following, we will consider the situation in which different outputs are produced concurrently, 
whereas inputs are heat and/or consuming work.
Quantifying and comparing the usefulness of
different outputs in response to the provided inputs has been discussed recently \cite{Manzano2020}. In there, the efficiency for a particular operating mode (consisting in a set of inputs and outputs) is given as 
\begin{equation}
\label{eq:efficiency}
    \eta^{\text{(mode)}} = \frac{\sum^+_{\alpha}\dot{w}^\alpha+\sum_{j}^+J_j\left(\frac{T_{\rm r}}{T_{j}}-1\right) }{-\sum_\alpha^-\dot{w}^{\alpha}-\sum_{j}^-J_j\left(\frac{T_{\rm r}}{T_{j}}-1\right)}\leq 1.
\end{equation}
where modes of hybrid machine $m \in \{P,R,E,ER,EP,RP\}$. Here, $\sum^{\pm}_j J_j = \sum_j (J_j\pm |J_j|)/2$ are the sums over the positive and negative heat currents $J_j$ and $\dot{w}$ is the production rate of work performed by the system~\footnote{
The sign convention is adopted such that a current is positive when it flows to the system. For instance, when a system is in the RP mode, it pumps heat to a hot reservoir, $J_h<0$, and it refrigerates a cold reservoir, $J_c>0$, by the work provided from the environment, $\dot w<0$. In this case, the denominator of the efficiency expression is $-\sum_\alpha^-\dot{w}^{\alpha}$, and the numerator is $J_j\left(\frac{T_{\rm r}}{T_{j}}-1\right)$. That is, currents doing useful works are put in the numerator, while a current providing a resource is put in the denominator.
}. The choice of the reference temperature $T_{\rm r}$ depends on the multiterminal configuration and the input/output tasks.
Let us focus on the case of three terminals. This scenario applies to our setup  since although the two quantum dot setup in Fig.1(a) contains the four leads, there is no particle and energy flow to the superconducting (SC) lead in the large pairing gap limit. The SC lead is analytically integrated in our treatment. (extension to n-baths will be given elsewhere). The MTUR reads ($k_B=1$)~\cite{dechant2018multidimensional}
\begin{equation} \label{eq:tur_0}
    \bold J_0 ^T D^{-1} \bold J_0 \leq \frac 1 2 \sigma,
\end{equation}
where  vector $\bold J_0 =(J_r,J_h,J_c)^T$ is a three component vector with heat current via reference lead, hot lead, and cold lead, respectively. The matrix element $D_{ij}$ contains the physical meaning of thermal conductance between lead $i$ and $j$ according to the fluctuation-dissipation theorem. 
\
The implication of the relation is that the precision of measurement ($ \bold J_0 ^T D^{-1} \bold J_0  \equiv 1/\Delta P $) is upper bounded by the entropy production $\sigma$. That is, the uncertainty of measurement precision $\Delta P$
can be enhanced at the expense of entropy production, $ \sigma\,\Delta P \geq 2$. 

To make a connection with the thermal efficiency, which is in terms of heat current and the projection rate of work, we perform a transformation to $\bold J = (J_w,J_h, J_c)^T= {\mathcal A} \bold J_0$ using  $J_w = \dot w = J_r + J_h + J_c= - \sum_{j=r,h,c} \mu_j I_j$, where $\mathcal A$ is a 3$\times$3 matrix 
\footnote{Explicitly,
\begin{eqnarray}
    \bold J &= \left(\begin{array}{ccc}
        1 & 1 & 1 \\ 0 & 1 & 0 \\ 0 & 0 & 1
    \end{array} \right)\left(
    \begin{array}{c}
        J_r \\ J_h \\ J_c
    \end{array}\right) = \mathcal A \bold J_0.
\end{eqnarray}
}.
The entropy production rate is conveniently expressed by the current vector:  
\begin{eqnarray}
   \sigma &= \sum_{j=r,h,c} \frac{\dot Q_j}{T_j} = \frac{1}{T_r}\left( \bold {t_w} \cdot \bold J - \bold{t_u} \cdot \bold J \right) ,     
\end{eqnarray}
 where $\bold t_w$ and $\bold t_u$ are vectors projecting the current to the one generating and reducing the entropy, respectively. 
The former current $J_w=\hat {\bold t}_w \cdot \bold J$ consumes a resource  to generate useful current $J_u=\hat{\bold t}_u \cdot \bold J$. 
For instance, when a system is in the RP mode, $\bold t_w^{\text{RP}} = (-1,0,0)$ and $\bold t_u^{\text{RP}} = (0, - 1+\frac{T_r}{T_h} ,\frac{T_r}{T_c}-1 )$.  The machine receives work from the environment and then use the resource to pump heat and to refrigerate the cold reservoir. The efficiency is expressed as follows: 
\begin{eqnarray}
    \eta^{RP} = \frac{\bold t_u \cdot \bold J}{\bold t_w \cdot \bold J} = \eta^{R}+\eta^{P},
\end{eqnarray}
where the hybrid efficiency is then divided into the two separate single efficiencies. 
The other hybrid working modes can be similarly expressed by choosing different $\bold t_{u,w}$ vectors while maintaining the same vectorial expression for entropy production $\sigma$ and the efficiency $\eta^{m}$ shown above~\footnote {For instance, 
\begin{eqnarray}
    \bold t_w^{\text{EP}} =(0,0,-\frac{T_r}{T_c}+1), \,\,\,\, \bold t_u^{\text{EP}} = (1,-1 + \frac{T_r}{T_h}, 0), \\
    \bold t_w^{\text{ER}} = (0, 1-\frac{T_r}{T_h}, 0),\,\,\,\, \bold t_u^{\text{ER}} = (1, 0, \frac{T_r}{T_c}-1).
\end{eqnarray}
where the ER mode is generating work and refrigerating, and the EP mode is generating work and heat pumping.}. 
We introduce one more vector $\bold t^{\text{RP}}_\perp = (0, -\frac{T_r}{T_c}+1 , -1+\frac{T_r}{T_h})$ to be perpendicular to both $\bold t_{w,u}$. The expression of the entropy production and the efficiency motivate us to write the current vector in the basis of $\bold t_{u,w,\perp}$. 
$\bold J_t = (J_w,J_u,J_\perp)^T= \mathcal B \bold J,
$
where  $\mathcal B$ is a 3$\times$3 matrix in terms of temperature $T_{r,h,c}$ only
\footnote{
For the RP operating mode of a hybrid machine, $J_u= \left[ J_h (-1+\frac{T_r}{T_h}) + J_c (\frac{T_r}{T_c}-1) \right]/|\bold t_u|$ and $J_\perp= \left[-J_h (\frac{T_r}{T_h}-1) + J_c (-1 + \frac{T_r}{T_c}) \right]/|\bold t_\perp|$, where $|\bold t_u|=|\bold t_\perp |$. Hence, 
\begin{eqnarray}
\mathcal B = 
\left(\begin{array}{ccc}
        1 & 0 & 0 \\ 
    0 & \cos \alpha & \sin \alpha \\ 
    0 & -\sin \alpha & \cos \alpha 
    \end{array} \right), 
\end{eqnarray}
where $\cos \alpha = (-1+\frac{T_r}{T_h}) /|\bold t_u|$ and $\sin \alpha = (\frac{T_r}{T_c}-1) /|\bold t_u|$. 
}. 
The MTUR is then expressed in terms of new current vector: 
\begin{equation}
\bold J_t ^T  \tilde D^{-1} \bold J_t \leq\frac{1}{T_r}(\bold t_u \cdot \bold J) (1 - \eta ^{-1}), 
\end{equation}
where $\tilde D =  [\mathcal{B} \mathcal{A} D \mathcal{A}^T \mathcal{B}^T]$. The expression is utilized to obtain the upper bound of current of interest, for example, `useful' current, $J_u$, is upper bounded by 
\begin{eqnarray} \label{eq:ju}
    J_u \leq  \frac{1}{T_r}|\bold t_u| (1 - \eta ^{-1}) \left[\sum_{i,j=u,w,\perp} ( \tilde D^{-1} )_{ij} \frac{J_i J_j}{J_u^2}  \right]^{-1}.
\end{eqnarray}
where the ratio of currents $J_i/J_u$ is fixed once efficiency $\eta^{R,P}$ are given
~\footnote{In the RP mode, $\frac{J_w}{J_u} = -(\eta^{RP})^{-1} $, 
and $\frac{J_\perp}{J_u} = -\frac{1}{\eta^{RP}}\left[ \eta^P r_t - \eta^R r_t^{-1} \right] $, where $r_t = \frac{T_r/T_c - 1}{-1+T_r/T_h} $. Hence, the upper bound in Eq.~(\ref{eq:ju}) is determined by thermal conductance tensor, temperatures and efficiency.    }
. 
The upper bound of the thermal efficiency is given by the Clausius relation. For a given thermal efficiency, the MTUR provides the quantitative upper bound of useful current in terms of temperatures and thermal conductance tensor. 
That is practically relevant information when we need the generation of a certain amount of current regardless of an optimal thermal efficiency. For instance, at the optimal efficiency, it is well known that the amount of useful current that can be generated is zero.  When a system contains quantum elements that promote the quantum coherence in particle and energy transport, the upper bound of the MTUR can be exceeded, and as a result one can harness resource from a hybrid thermal machine beyond the classical upper bound. In the following we explicitly show in our double quantum dot setup the situation where the MTUR relation is violated as a result of the LDB breaking by Cooper pair transport. 


\begin{figure}[t]
\centering
\includegraphics[width=1\linewidth]{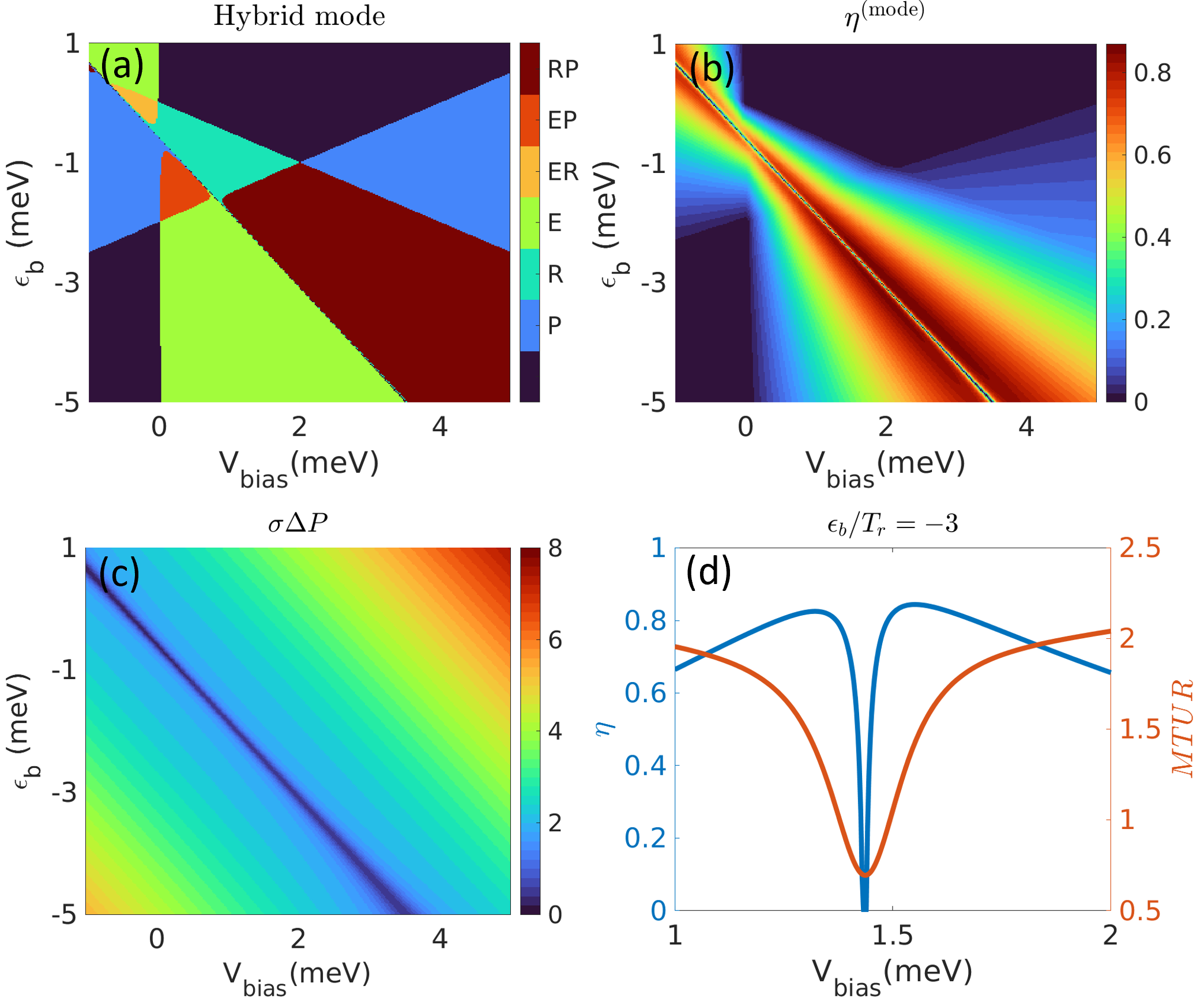}
\caption{
The hybrid thermal machine configured to show six working modes as indicated in (a) by colors in the domain of $(V_{\text{bias}}, \epsilon_b)$. $\mu_r=-0.04$, $T_h=1.4$, $T_c=0.6$, $\epsilon_t=4$, $U_t=1$, $U_{tb}=2$. $\Gamma_r=0.05$, $\Gamma_S=0.2$. The energies are indicated in unit of reference temperature, $T_r=1$. (b) Shows the efficiency of corresponding hybrid modes. (c) Illustrates the value of the MTUR, $\sigma\Delta P$. The regime with values less than 2 shows the violation. (d) Shows the region which shows the highest efficiency and the violation of MTUR simultaneously at $\epsilon_r/T_r=-3$. 
}
\label{fig:3}
\end{figure}

\emph{A superconducting hybrid machine---} To illustrate the validity and the departure of the MTUR due to quantum effects we consider parallel double quantum dots labelled by "t" (top) and "b" (bottom).
See Fig.~\ref{fig:1}(a) for the scheme of the setup. The top dot is connected to normal and superconducting leads and the intradot Coulomb interaction strength in this dot is denoted by $U_{t}$.
The bottom dot is attached to two normal leads with temperature $T_{h,c}$. 
The two dots are coupled capacitively and the interaction strength is $U_{tb}$. 
In the limit of  large gap (i.e., $\Delta\to\infty$), the superconducting lead can be easily integrated out~\cite{PhysRevLett.82.2788}.  Under this condition, the Hamiltonian  then reads $\calh = \calh_{\rm C}^{\rm eff} + \calh_{R} + \calh_T$ with \cite{Arovas00}
\begin{eqnarray}
\calh_{\rm C}^{\rm eff} &=& \epsilon_b d_b^{\dag}d_b + \sum_{\sm\in \{ \uparrow,\downarrow\}} \epsilon_{t\sm} d_{t\sm}^{\dag}d_{t\sm} + \Gamma_S \left(d_{t\up}^{\dag} d_{t\down}^{\dag} + \rm{H.c.}\right) \nonumber
           \\ &&+ U_{tb} d_b^{\dag}d_b \sum_{\sm \in \{ \uparrow,\downarrow\}} d_{t\sm}^{\dag}d_{t\sm} + U_t d_{t\up}^{\dag}d_{t\up}d_{t\down}^{\dag}d_{t\down}.
\label{eq:heff}
\end{eqnarray}
Here, $\calh_{R}$ is the Hamiltonian for the uncoupled  leads. The tunnelling from lead to dot is described by the Hamiltonian $\calh_{T}$.
The hybridization between lead and dot is given by $\Gamma_l$ with $l=\{r,h,c\}$.
The energy levels of the top and bottom dots are indicated by $\epsilon_{t\sm}$ and $\epsilon_b$, respectively.
The operators $d_{t\sm}^{\dag}$ ($d_{t\sm}$) and $d_{b}^{\dag}$ ($d_{b}$) stand for the creation (annihilation) operators for electrons in the top and bottom dots with $\sigma=\{\uparrow,\downarrow\}$. 
Hereafter, we deploy our formalism using the master equation approach. 
With this aim we  diagonalize the Hamiltonian given by Eq.~(\ref{eq:heff}).
To do so, we take $|n_b,n_t\rangle$ as the basis,
where $n_b = \{0,1\}$ is the occupation for the spinless bottom dot and $n_t = \{0,\uparrow/\downarrow,2\}$ the corresponding 
occupation for the top dot. When we diagonalize Eq.~(\ref{eq:heff}) in this basis, the hybridization $\Gamma_S$ between the top 
dot and the superconducting lead mixes coherently states with even number of electrons. The eigenstates are then given by $|n_b,
\uparrow/\downarrow\rangle$, and
$|n_b,\pm\rangle = u_{n_b,\pm} |n_b,0\rangle+ v_{n_b,\pm}|n_b,2\rangle$, where $u_{n_b,\pm}= (\epsilon_{n_b} \pm \sqrt{\epsilon_{n_b}^2+
\Gamma_S^2})/\mathcal{N}_{n_b,\pm}$  and $v_{n_b,\pm}=-\Gamma_S/\mathcal{N}_{n_b,\pm}$ with $\epsilon_{n_b} 
=\epsilon_t + U_t/2 + n_b U_{tb}$. Here, $\mathcal{N}_{n_b,\pm}$ is the normalization factor. The eigen energies for $|n_b,\pm\rangle$ are $E_{\ket{n_b\pm}} = n_b\epsilon_b + \epsilon_{n_b} \mp \sqrt{\epsilon_{n_b}^2 + \Gamma_S^2}$
whereas for $|n_b,\sm\rangle$, these are simply $E_{\ket{n_b\sigma} }=\epsilon_t+n_b (\epsilon_b+ U_{tb})$.

\emph{Pauli master equation---} We apply the master equation formalism to describe the dynamics for the occupation probability $\rho_{\theta}$ of the eigenstate $\ket{\theta}$. To this end, we write down the transition rate from the eigenstate $|\theta\rangle$ to $|\zeta\rangle$ as through the $\ell$ barrier across the $X=t,b$ quantum dot
\begin{equation}
\gamma_{l |\zeta\rangle \leftarrow |\theta\rangle }^{e(h)} = \Gamma_{l} \left|\langle \zeta |\Delta^{e(h)}_X|\theta\rangle\right |^2  
f^{e(h)}_{l}(E_{\theta}-E_{\zeta}),
\end{equation}
where $\Delta^{e(h)}_t=\sum_{\sigma} d_{t\sm}^\dagger (d_{t\sm})$ and for $\Delta^{e(h)}_b=d^\dagger(d)$.
In the limit $k_BT_l \gg \Gamma_l$, the  dynamics of the system is governed by the sequential tunneling events and we safely neglect higher-order tunneling correlations.  As we are interested in the calculation of the current fluctuations we write down the generalize master equation equation considering the counting fields that generate the complete full counting statistics $d\rho(\chi)/dt = -\call\rho(\chi$)
where $\call$ means Louvillian matrix. In the stationary limit  $\call \rho =0$, we obtain the full counting statistics formalism described in the supplemental material~\cite{supp}. 

\begin{figure}[t]
\centering
\includegraphics[width=1\linewidth]{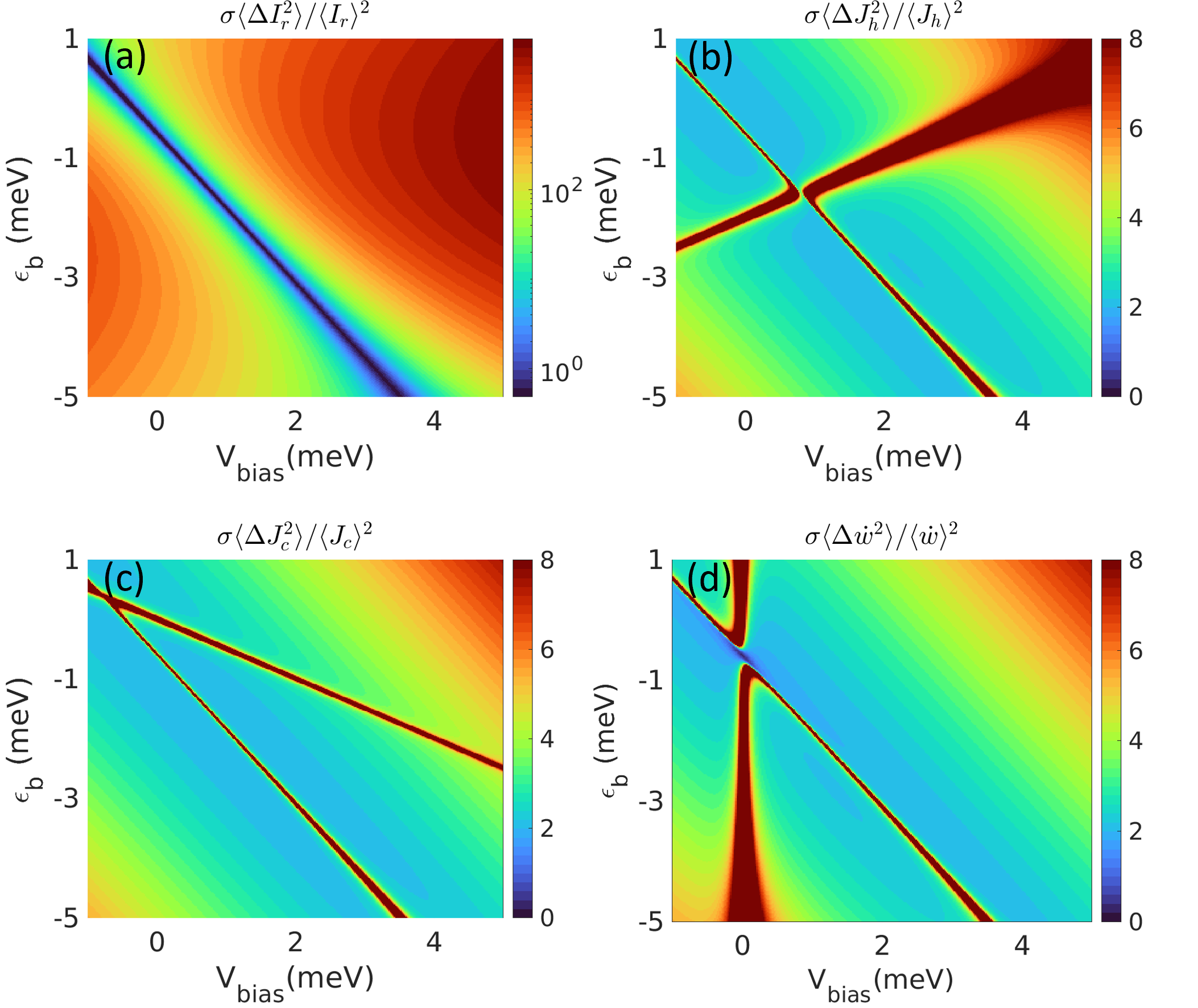}
\caption{The quantity from which the violation of the TUR can be read off is plotted for individual lead. (a) $\sigma \langle \Delta I_r^2 \rangle/\langle I_r \rangle^2$, associated with the charge current from the reference lead. 
(b,c) $\sigma \langle \Delta J_{h,c}^2 \rangle/\langle J_{h,c} \rangle^2$, heat current from the hot and cold lead, respectively. (d) $\sigma \langle \Delta \dot w^2 \rangle/\langle \dot w \rangle^2$, associated with the production rate of work (d). The transition between thermal working modes is reflected in the divergence of the TUR value plotted. The breaking of the local detailed balance appears in the particle current from the reference lead due to the proximity effect from superconducting lead.}
\label{fig:4}
\end{figure}
\emph{Breakdown of Local Detailed Balance---}
Local detailed balance \cite{PhysRevB.28.1655} amounts to requiring that the log-ratio of an individual transition to the probability of its time-reversed transition equals the entropy flux: $\gamma_F/\gamma_B=e^{-\sigma_C}$. In systems where LDB is broken exhibits different entropy productions when they are calculated by following the Clausius relation ($\sigma_C$) and the Shannon entropy expression ($\sigma_S$). Due to the breakdown of the LDB an information flow ($I_F$) is established by doing $\sigma_S=\sigma_C+I_F$ from the Maxwell demon point of view \cite{Esposito_2012,Guillem2017}. In our device LDB is manifestly broken having two consequences, namely (i) the appearance of drag currents as reported in Ref. \cite{Tabatabaei2020}, and (ii) the violation of the thermodynamic uncertainty relation \cite{Christian2021} which is our main focus. Due to the Andreev processes, the forward and backward transition processes are not connected by Boltzmann weights. 
The ratio of the forward ($\ket{n_b,\sigma}\rightarrow \ket{n_b,\pm}$) to backward ($\ket{n_b,\pm}\rightarrow \ket{n_b,\sigma}$) transition is (See \cite{supp} for full details)

\begin{equation}
\label{ldb}
    \frac{\gamma_B}{\gamma_F}=e^{\frac{\Delta E}{T_r}}\left[ \frac{\cosh\theta_\gamma +e^{-\frac{\Delta E}{T_r}}\cosh (\theta_\gamma-\theta_{\mu_r})}{\cosh(\theta_\gamma+\theta_{\mu_r}) +e^{-\frac{\Delta E}{T_r}}\cosh \theta_\gamma} \right],
\end{equation}
where $\Delta E=E_{\ket{n_b\pm}}-E_{\ket{n_b\sm}}$, $\cosh \theta_\gamma = \frac{1}{2}(\left|\frac{v_{n_b,\pm}}{u_{n_b,\pm}} \right|^2+ \left|\frac{u_{n_b,\pm}}{v_{n_b,\pm}} \right|^2)$ and $\cosh\theta_{\mu_r}=\frac{1}{2}(e^{\frac{\mu_r}{T_r}} + e^{-\frac{\mu_r}{T_r}} )$.  $\mu_r$ is the chemical potential of reference lead. 
The ratio in Eq. (\ref{ldb})  depends on the the superconducting quasiparticle amplitudes $u_{n_b\pm}/v_{n_b\pm}$ when $\mu_r\neq 0$, thus the LDB is broken. The LDB is restored when the proximity effect in the top quantum dot is removed, $\theta_{\gamma} \rightarrow \infty$. 
In brief, Andreev coherent processes are responsible for the breakdown of the LDB under nonequilibrium conditions being the main cause for the deviation of the TUR. 
\newline

\emph{Results: RP, EP, and ER modes---}
 The versatile double quantum dot system shows the all possible hybrid machine modes as indicated in Fig.~\ref{fig:3}(a) by tuning $\epsilon_b$ and $V_{\text{bias}}=\mu_h-\mu_c$.
The hybrid machine efficiency is achieved as high as $\eta \sim 0.85$ closely following the boundary where the three machine modes (E, P and R) are present nearby each other. For instance, the boundary of E and RP, that of R and EP, and that of P and ER is the close neighbor of the high efficiency region shown in Fig.~\ref{fig:3}(b). Note that the thermal efficiency expression is dependent of the modes, Eq.~(\ref{eq:efficiency}). When the chemical potential of the reference lead $\mu_r$ is set nonzero, the MTUR relation is violated at those boundaries of three machine modes as shown in $\sigma\Delta P$ plotted in Fig.~\ref{fig:3}(c). This implies that it is possible to achieve the optimally operating thermal machine generating useful currents exceeding the upper bound limitation set by the classical MTUR. The violation is associated with the quantum transport mainly involved in the upper quantum within the proximity effect of superconducting lead. The introduction of $\mu_r\neq 0$ triggers the quantum effect kicked in the machine operation. The claim is supported in Fig.~\ref{fig:4}(a), where the TUR associated with the charge current via the reference lead shows the violation,  $\sigma(\Delta P)_{r,part} \equiv \sigma \langle\Delta I_r^2\rangle/I_r^2 < 2$.

The six hybrid machine modes are determined based on the sign of elements of current vector $\bold J=(\dot w,J_h,J_c)$. The mode of thermal machine  switches when a current element approaches to zero and then flips its sign. In the course, the uncertainty of  measurement precision $\Delta P_i \equiv \langle \Delta J_i^2 \rangle/J_i^2, i\in \{w,h,c \}$ diverges. Figure \ref{fig:4}(b,c,d) shows $\sigma \Delta P_{i=h,c,w}$ associated with the current of immediate interest, $J_{h,c,w}$. The phase boundary of the thermal machine modes $P,R,E$ can be read off from the divergence of $\sigma \Delta P_{i=h,c,w}$ which is consistent with the diagram of  modes drawn in  Fig.~\ref{fig:3}(a). It is worth noting that the TUR of heat current associated with single lead, $\sigma \Delta P_{i=r,h,c}$, does not show the violation, while the MTUR of heat current drawn in Fig.~\ref{fig:3}(c) does. It implies that in a multiterminal system the information contained in the correlation of cross terminal currents is crucial to properly characterize the system.

{\it Conclusion---}
\
The thermodynamic uncertainty relation has been formulated for multiple-bath devices and it establishes a trade-off between generalized current fluctuations and entropy production. A new class of machines, the so-called hybrid machines,  are devices that are able to perform several useful tasks at once. Here, we have applied the uncertainty relation to deal with multiterminal devices working as hybrid machines. From the MTUR a new bound is established for the resourceful current in terms of the generalized efficiency and the cross current correlations. We have shown that  MTUR are violated in quantum systems as a result of the LDB breaking. 
We have shown that depending on the operating mode (RP, EP, or ER) a violation of the MTUR is accompanied by a high efficiency being the optimal scenario for an hybrid thermal machine. We provide a general guideline for the design of multiterminal hybrid machines in the quantum regime that operate at the best task performance, meaning low dissipation, small fluctuations around average values and high task efficiency with a moderately high resourceful current. 

{\it Acknowledgements---}
R.L acknowledges the financial support through the grants PID2020-117347GB-I00 and the grant from the María de Maeztu Program for Units of Excellence No. MDM2017-0711 funded by MCIN/AEI/10.13039/501100011033'. K.W.K.~acknowledges financial support by Basic Science Research Program through the National Research Foundation of Korea (NRF) funded by the Ministry of Education (20211060) and Korea government(MSIT) (No.2020R1A5A1016518).

\bibliography{bibliography}

 \newcommand{\noop}[1]{}
\begin{thebibliography}{53}%
\makeatletter
\providecommand \@ifxundefined [1]{%
 \@ifx{#1\undefined}
}%
\providecommand \@ifnum [1]{%
 \ifnum #1\expandafter \@firstoftwo
 \else \expandafter \@secondoftwo
 \fi
}%
\providecommand \@ifx [1]{%
 \ifx #1\expandafter \@firstoftwo
 \else \expandafter \@secondoftwo
 \fi
}%
\providecommand \natexlab [1]{#1}%
\providecommand \enquote  [1]{``#1''}%
\providecommand \bibnamefont  [1]{#1}%
\providecommand \bibfnamefont [1]{#1}%
\providecommand \citenamefont [1]{#1}%
\providecommand \href@noop [0]{\@secondoftwo}%
\providecommand \href [0]{\begingroup \@sanitize@url \@href}%
\providecommand \@href[1]{\@@startlink{#1}\@@href}%
\providecommand \@@href[1]{\endgroup#1\@@endlink}%
\providecommand \@sanitize@url [0]{\catcode `\\12\catcode `\$12\catcode
  `\&12\catcode `\#12\catcode `\^12\catcode `\_12\catcode `\%12\relax}%
\providecommand \@@startlink[1]{}%
\providecommand \@@endlink[0]{}%
\providecommand \url  [0]{\begingroup\@sanitize@url \@url }%
\providecommand \@url [1]{\endgroup\@href {#1}{\urlprefix }}%
\providecommand \urlprefix  [0]{URL }%
\providecommand \Eprint [0]{\href }%
\providecommand \doibase [0]{https://doi.org/}%
\providecommand \selectlanguage [0]{\@gobble}%
\providecommand \bibinfo  [0]{\@secondoftwo}%
\providecommand \bibfield  [0]{\@secondoftwo}%
\providecommand \translation [1]{[#1]}%
\providecommand \BibitemOpen [0]{}%
\providecommand \bibitemStop [0]{}%
\providecommand \bibitemNoStop [0]{.\EOS\space}%
\providecommand \EOS [0]{\spacefactor3000\relax}%
\providecommand \BibitemShut  [1]{\csname bibitem#1\endcsname}%
\let\auto@bib@innerbib\@empty
\bibitem [{\citenamefont {Carnot}(1978)}]{Carnot}%
  \BibitemOpen
  \bibfield  {author} {\bibinfo {author} {\bibfnamefont {S.}~\bibnamefont
  {Carnot}},\ }\href@noop {} {\emph {\bibinfo {title} {R{\'e}flexions sur la
  puissance motrice du feu}}},\ \bibinfo {number} {26}\ (\bibinfo  {publisher}
  {Vrin},\ \bibinfo {year} {1978})\BibitemShut {NoStop}%
\bibitem [{\citenamefont {Barato}\ and\ \citenamefont
  {Seifert}(2015)}]{Barato2015}%
  \BibitemOpen
  \bibfield  {author} {\bibinfo {author} {\bibfnamefont {A.~C.}\ \bibnamefont
  {Barato}}\ and\ \bibinfo {author} {\bibfnamefont {U.}~\bibnamefont
  {Seifert}},\ }\bibfield  {title} {\bibinfo {title} {Thermodynamic uncertainty
  relation for biomolecular processes},\ }\href@noop {} {\bibfield  {journal}
  {\bibinfo  {journal} {Physical review letters}\ }\textbf {\bibinfo {volume}
  {114}},\ \bibinfo {pages} {158101} (\bibinfo {year} {2015})}\BibitemShut
  {NoStop}%
\bibitem [{\citenamefont {Gingrich}\ \emph {et~al.}(2016)\citenamefont
  {Gingrich}, \citenamefont {Horowitz}, \citenamefont {Perunov},\ and\
  \citenamefont {England}}]{Todd2016}%
  \BibitemOpen
  \bibfield  {author} {\bibinfo {author} {\bibfnamefont {T.~R.}\ \bibnamefont
  {Gingrich}}, \bibinfo {author} {\bibfnamefont {J.~M.}\ \bibnamefont
  {Horowitz}}, \bibinfo {author} {\bibfnamefont {N.}~\bibnamefont {Perunov}},\
  and\ \bibinfo {author} {\bibfnamefont {J.~L.}\ \bibnamefont {England}},\
  }\bibfield  {title} {\bibinfo {title} {Dissipation bounds all steady-state
  current fluctuations},\ }\href@noop {} {\bibfield  {journal} {\bibinfo
  {journal} {Physical review letters}\ }\textbf {\bibinfo {volume} {116}},\
  \bibinfo {pages} {120601} (\bibinfo {year} {2016})}\BibitemShut {NoStop}%
\bibitem [{\citenamefont {Pietzonka}\ \emph
  {et~al.}(2016{\natexlab{a}})\citenamefont {Pietzonka}, \citenamefont
  {Barato},\ and\ \citenamefont {Seifert}}]{Pietz2016}%
  \BibitemOpen
  \bibfield  {author} {\bibinfo {author} {\bibfnamefont {P.}~\bibnamefont
  {Pietzonka}}, \bibinfo {author} {\bibfnamefont {A.~C.}\ \bibnamefont
  {Barato}},\ and\ \bibinfo {author} {\bibfnamefont {U.}~\bibnamefont
  {Seifert}},\ }\bibfield  {title} {\bibinfo {title} {Universal bounds on
  current fluctuations},\ }\href@noop {} {\bibfield  {journal} {\bibinfo
  {journal} {Physical Review E}\ }\textbf {\bibinfo {volume} {93}},\ \bibinfo
  {pages} {052145} (\bibinfo {year} {2016}{\natexlab{a}})}\BibitemShut
  {NoStop}%
\bibitem [{\citenamefont {Pietzonka}\ \emph
  {et~al.}(2016{\natexlab{b}})\citenamefont {Pietzonka}, \citenamefont
  {Barato},\ and\ \citenamefont {Seifert}}]{Pietzonka16b}%
  \BibitemOpen
  \bibfield  {author} {\bibinfo {author} {\bibfnamefont {P.}~\bibnamefont
  {Pietzonka}}, \bibinfo {author} {\bibfnamefont {A.~C.}\ \bibnamefont
  {Barato}},\ and\ \bibinfo {author} {\bibfnamefont {U.}~\bibnamefont
  {Seifert}},\ }\bibfield  {title} {\bibinfo {title} {Universal bound on the
  efficiency of molecular motors},\ }\href@noop {} {\bibfield  {journal}
  {\bibinfo  {journal} {Journal of Statistical Mechanics: Theory and
  Experiment}\ }\textbf {\bibinfo {volume} {2016}},\ \bibinfo {pages} {124004}
  (\bibinfo {year} {2016}{\natexlab{b}})}\BibitemShut {NoStop}%
\bibitem [{\citenamefont {Pietzonka}\ and\ \citenamefont
  {Seifert}(2018)}]{Pietz2018}%
  \BibitemOpen
  \bibfield  {author} {\bibinfo {author} {\bibfnamefont {P.}~\bibnamefont
  {Pietzonka}}\ and\ \bibinfo {author} {\bibfnamefont {U.}~\bibnamefont
  {Seifert}},\ }\bibfield  {title} {\bibinfo {title} {Universal trade-off
  between power, efficiency, and constancy in steady-state heat engines},\
  }\href@noop {} {\bibfield  {journal} {\bibinfo  {journal} {Physical review
  letters}\ }\textbf {\bibinfo {volume} {120}},\ \bibinfo {pages} {190602}
  (\bibinfo {year} {2018})}\BibitemShut {NoStop}%
\bibitem [{\citenamefont {Gingrich}\ and\ \citenamefont
  {Horowitz}(2017)}]{Gingrich17}%
  \BibitemOpen
  \bibfield  {author} {\bibinfo {author} {\bibfnamefont {T.~R.}\ \bibnamefont
  {Gingrich}}\ and\ \bibinfo {author} {\bibfnamefont {J.~M.}\ \bibnamefont
  {Horowitz}},\ }\bibfield  {title} {\bibinfo {title} {Fundamental bounds on
  first passage time fluctuations for currents},\ }\href@noop {} {\bibfield
  {journal} {\bibinfo  {journal} {Physical review letters}\ }\textbf {\bibinfo
  {volume} {119}},\ \bibinfo {pages} {170601} (\bibinfo {year}
  {2017})}\BibitemShut {NoStop}%
\bibitem [{\citenamefont {Timpanaro}\ \emph {et~al.}(2019)\citenamefont
  {Timpanaro}, \citenamefont {Guarnieri}, \citenamefont {Goold},\ and\
  \citenamefont {Landi}}]{Timp2019}%
  \BibitemOpen
  \bibfield  {author} {\bibinfo {author} {\bibfnamefont {A.~M.}\ \bibnamefont
  {Timpanaro}}, \bibinfo {author} {\bibfnamefont {G.}~\bibnamefont
  {Guarnieri}}, \bibinfo {author} {\bibfnamefont {J.}~\bibnamefont {Goold}},\
  and\ \bibinfo {author} {\bibfnamefont {G.~T.}\ \bibnamefont {Landi}},\
  }\bibfield  {title} {\bibinfo {title} {Thermodynamic uncertainty relations
  from exchange fluctuation theorems},\ }\href@noop {} {\bibfield  {journal}
  {\bibinfo  {journal} {Physical review letters}\ }\textbf {\bibinfo {volume}
  {123}},\ \bibinfo {pages} {090604} (\bibinfo {year} {2019})}\BibitemShut
  {NoStop}%
\bibitem [{\citenamefont {Hasegawa}\ and\ \citenamefont
  {Van~Vu}(2019)}]{Hase2019}%
  \BibitemOpen
  \bibfield  {author} {\bibinfo {author} {\bibfnamefont {Y.}~\bibnamefont
  {Hasegawa}}\ and\ \bibinfo {author} {\bibfnamefont {T.}~\bibnamefont
  {Van~Vu}},\ }\bibfield  {title} {\bibinfo {title} {Fluctuation theorem
  uncertainty relation},\ }\href@noop {} {\bibfield  {journal} {\bibinfo
  {journal} {Physical review letters}\ }\textbf {\bibinfo {volume} {123}},\
  \bibinfo {pages} {110602} (\bibinfo {year} {2019})}\BibitemShut {NoStop}%
\bibitem [{\citenamefont {Pietzonka}\ \emph {et~al.}(2017)\citenamefont
  {Pietzonka}, \citenamefont {Ritort},\ and\ \citenamefont
  {Seifert}}]{Pietzonka17}%
  \BibitemOpen
  \bibfield  {author} {\bibinfo {author} {\bibfnamefont {P.}~\bibnamefont
  {Pietzonka}}, \bibinfo {author} {\bibfnamefont {F.}~\bibnamefont {Ritort}},\
  and\ \bibinfo {author} {\bibfnamefont {U.}~\bibnamefont {Seifert}},\
  }\bibfield  {title} {\bibinfo {title} {Finite-time generalization of the
  thermodynamic uncertainty relation},\ }\href@noop {} {\bibfield  {journal}
  {\bibinfo  {journal} {Physical Review E}\ }\textbf {\bibinfo {volume} {96}},\
  \bibinfo {pages} {012101} (\bibinfo {year} {2017})}\BibitemShut {NoStop}%
\bibitem [{\citenamefont {Horowitz}\ and\ \citenamefont
  {Gingrich}(2017)}]{Horowitz17}%
  \BibitemOpen
  \bibfield  {author} {\bibinfo {author} {\bibfnamefont {J.~M.}\ \bibnamefont
  {Horowitz}}\ and\ \bibinfo {author} {\bibfnamefont {T.~R.}\ \bibnamefont
  {Gingrich}},\ }\bibfield  {title} {\bibinfo {title} {Proof of the finite-time
  thermodynamic uncertainty relation for steady-state currents},\ }\href@noop
  {} {\bibfield  {journal} {\bibinfo  {journal} {Physical Review E}\ }\textbf
  {\bibinfo {volume} {96}},\ \bibinfo {pages} {020103} (\bibinfo {year}
  {2017})}\BibitemShut {NoStop}%
\bibitem [{\citenamefont {Shiraishi}(2017)}]{Shiraishi17}%
  \BibitemOpen
  \bibfield  {author} {\bibinfo {author} {\bibfnamefont {N.}~\bibnamefont
  {Shiraishi}},\ }\bibfield  {title} {\bibinfo {title} {Finite-time
  thermodynamic uncertainty relation do not hold for discrete-time markov
  process},\ }\href@noop {} {\bibfield  {journal} {\bibinfo  {journal} {arXiv
  preprint arXiv:1706.00892}\ } (\bibinfo {year} {2017})}\BibitemShut {NoStop}%
\bibitem [{\citenamefont {Proesmans}\ and\ \citenamefont {Van~den
  Broeck}(2017)}]{Proesmans17}%
  \BibitemOpen
  \bibfield  {author} {\bibinfo {author} {\bibfnamefont {K.}~\bibnamefont
  {Proesmans}}\ and\ \bibinfo {author} {\bibfnamefont {C.}~\bibnamefont
  {Van~den Broeck}},\ }\bibfield  {title} {\bibinfo {title} {Discrete-time
  thermodynamic uncertainty relation},\ }\href@noop {} {\bibfield  {journal}
  {\bibinfo  {journal} {EPL (Europhysics Letters)}\ }\textbf {\bibinfo {volume}
  {119}},\ \bibinfo {pages} {20001} (\bibinfo {year} {2017})}\BibitemShut
  {NoStop}%
\bibitem [{\citenamefont {Macieszczak}\ \emph {et~al.}(2018)\citenamefont
  {Macieszczak}, \citenamefont {Brandner},\ and\ \citenamefont
  {Garrahan}}]{Macieszczak18}%
  \BibitemOpen
  \bibfield  {author} {\bibinfo {author} {\bibfnamefont {K.}~\bibnamefont
  {Macieszczak}}, \bibinfo {author} {\bibfnamefont {K.}~\bibnamefont
  {Brandner}},\ and\ \bibinfo {author} {\bibfnamefont {J.~P.}\ \bibnamefont
  {Garrahan}},\ }\bibfield  {title} {\bibinfo {title} {Unified thermodynamic
  uncertainty relations in linear response},\ }\href@noop {} {\bibfield
  {journal} {\bibinfo  {journal} {Physical review letters}\ }\textbf {\bibinfo
  {volume} {121}},\ \bibinfo {pages} {130601} (\bibinfo {year}
  {2018})}\BibitemShut {NoStop}%
\bibitem [{\citenamefont {Agarwalla}\ and\ \citenamefont
  {Segal}(2018{\natexlab{a}})}]{Bija2018}%
  \BibitemOpen
  \bibfield  {author} {\bibinfo {author} {\bibfnamefont {B.~K.}\ \bibnamefont
  {Agarwalla}}\ and\ \bibinfo {author} {\bibfnamefont {D.}~\bibnamefont
  {Segal}},\ }\bibfield  {title} {\bibinfo {title} {Assessing the validity of
  the thermodynamic uncertainty relation in quantum systems},\ }\href@noop {}
  {\bibfield  {journal} {\bibinfo  {journal} {Physical Review B}\ }\textbf
  {\bibinfo {volume} {98}},\ \bibinfo {pages} {155438} (\bibinfo {year}
  {2018}{\natexlab{a}})}\BibitemShut {NoStop}%
\bibitem [{\citenamefont {Marsland}\ and\ \citenamefont
  {England}(2017)}]{Mars2018}%
  \BibitemOpen
  \bibfield  {author} {\bibinfo {author} {\bibfnamefont {R.}~\bibnamefont
  {Marsland}}\ and\ \bibinfo {author} {\bibfnamefont {J.}~\bibnamefont
  {England}},\ }\bibfield  {title} {\bibinfo {title} {Limits of predictions in
  thermodynamic systems: a review},\ }\href@noop {} {\bibfield  {journal}
  {\bibinfo  {journal} {Reports on Progress in Physics}\ }\textbf {\bibinfo
  {volume} {81}},\ \bibinfo {pages} {016601} (\bibinfo {year}
  {2017})}\BibitemShut {NoStop}%
\bibitem [{\citenamefont {Potts}\ and\ \citenamefont
  {Samuelsson}(2019)}]{Potts19}%
  \BibitemOpen
  \bibfield  {author} {\bibinfo {author} {\bibfnamefont {P.~P.}\ \bibnamefont
  {Potts}}\ and\ \bibinfo {author} {\bibfnamefont {P.}~\bibnamefont
  {Samuelsson}},\ }\bibfield  {title} {\bibinfo {title} {Thermodynamic
  uncertainty relations including measurement and feedback},\ }\href@noop {}
  {\bibfield  {journal} {\bibinfo  {journal} {Physical Review E}\ }\textbf
  {\bibinfo {volume} {100}},\ \bibinfo {pages} {052137} (\bibinfo {year}
  {2019})}\BibitemShut {NoStop}%
\bibitem [{\citenamefont {Manzano}\ \emph
  {et~al.}(2020{\natexlab{a}})\citenamefont {Manzano}, \citenamefont
  {S\'anchez}, \citenamefont {Silva}, \citenamefont {Haack}, \citenamefont
  {Brask}, \citenamefont {Brunner},\ and\ \citenamefont {Potts}}]{Gonzalo2020}%
  \BibitemOpen
  \bibfield  {author} {\bibinfo {author} {\bibfnamefont {G.}~\bibnamefont
  {Manzano}}, \bibinfo {author} {\bibfnamefont {R.}~\bibnamefont {S\'anchez}},
  \bibinfo {author} {\bibfnamefont {R.}~\bibnamefont {Silva}}, \bibinfo
  {author} {\bibfnamefont {G.}~\bibnamefont {Haack}}, \bibinfo {author}
  {\bibfnamefont {J.~B.}\ \bibnamefont {Brask}}, \bibinfo {author}
  {\bibfnamefont {N.}~\bibnamefont {Brunner}},\ and\ \bibinfo {author}
  {\bibfnamefont {P.~P.}\ \bibnamefont {Potts}},\ }\bibfield  {title} {\bibinfo
  {title} {Hybrid thermal machines: Generalized thermodynamic resources for
  multitasking},\ }\href@noop {} {\bibfield  {journal} {\bibinfo  {journal}
  {Phys. Rev. Research}\ }\textbf {\bibinfo {volume} {2}},\ \bibinfo {pages}
  {043302} (\bibinfo {year} {2020}{\natexlab{a}})}\BibitemShut {NoStop}%
\bibitem [{\citenamefont {L\'opez}\ \emph {et~al.}(2012)\citenamefont
  {L\'opez}, \citenamefont {Lim},\ and\ \citenamefont {S\'anchez}}]{SFR}%
  \BibitemOpen
  \bibfield  {author} {\bibinfo {author} {\bibfnamefont {R.}~\bibnamefont
  {L\'opez}}, \bibinfo {author} {\bibfnamefont {J.~S.}\ \bibnamefont {Lim}},\
  and\ \bibinfo {author} {\bibfnamefont {D.}~\bibnamefont {S\'anchez}},\
  }\bibfield  {title} {\bibinfo {title} {Fluctuation relations for
  spintronics},\ }\href {https://doi.org/10.1103/PhysRevLett.108.246603}
  {\bibfield  {journal} {\bibinfo  {journal} {Phys. Rev. Lett.}\ }\textbf
  {\bibinfo {volume} {108}},\ \bibinfo {pages} {246603} (\bibinfo {year}
  {2012})}\BibitemShut {NoStop}%
\bibitem [{\citenamefont {Ptaszy{\'n}ski}(2018)}]{Ptasz2018}%
  \BibitemOpen
  \bibfield  {author} {\bibinfo {author} {\bibfnamefont {K.}~\bibnamefont
  {Ptaszy{\'n}ski}},\ }\bibfield  {title} {\bibinfo {title} {Coherence-enhanced
  constancy of a quantum thermoelectric generator},\ }\href@noop {} {\bibfield
  {journal} {\bibinfo  {journal} {Physical Review B}\ }\textbf {\bibinfo
  {volume} {98}},\ \bibinfo {pages} {085425} (\bibinfo {year}
  {2018})}\BibitemShut {NoStop}%
\bibitem [{\citenamefont {Agarwalla}\ and\ \citenamefont
  {Segal}(2018{\natexlab{b}})}]{Agar2018}%
  \BibitemOpen
  \bibfield  {author} {\bibinfo {author} {\bibfnamefont {B.~K.}\ \bibnamefont
  {Agarwalla}}\ and\ \bibinfo {author} {\bibfnamefont {D.}~\bibnamefont
  {Segal}},\ }\bibfield  {title} {\bibinfo {title} {Assessing the validity of
  the thermodynamic uncertainty relation in quantum systems},\ }\href@noop {}
  {\bibfield  {journal} {\bibinfo  {journal} {Physical Review B}\ }\textbf
  {\bibinfo {volume} {98}},\ \bibinfo {pages} {155438} (\bibinfo {year}
  {2018}{\natexlab{b}})}\BibitemShut {NoStop}%
\bibitem [{\citenamefont {Brandner}\ \emph
  {et~al.}(2018{\natexlab{a}})\citenamefont {Brandner}, \citenamefont
  {Hanazato},\ and\ \citenamefont {Saito}}]{Brand2018}%
  \BibitemOpen
  \bibfield  {author} {\bibinfo {author} {\bibfnamefont {K.}~\bibnamefont
  {Brandner}}, \bibinfo {author} {\bibfnamefont {T.}~\bibnamefont {Hanazato}},\
  and\ \bibinfo {author} {\bibfnamefont {K.}~\bibnamefont {Saito}},\ }\bibfield
   {title} {\bibinfo {title} {Thermodynamic bounds on precision in ballistic
  multiterminal transport},\ }\href@noop {} {\bibfield  {journal} {\bibinfo
  {journal} {Physical review letters}\ }\textbf {\bibinfo {volume} {120}},\
  \bibinfo {pages} {090601} (\bibinfo {year} {2018}{\natexlab{a}})}\BibitemShut
  {NoStop}%
\bibitem [{\citenamefont {Liu}\ and\ \citenamefont {Segal}(2019)}]{Liu2019}%
  \BibitemOpen
  \bibfield  {author} {\bibinfo {author} {\bibfnamefont {J.}~\bibnamefont
  {Liu}}\ and\ \bibinfo {author} {\bibfnamefont {D.}~\bibnamefont {Segal}},\
  }\bibfield  {title} {\bibinfo {title} {Thermodynamic uncertainty relation in
  quantum thermoelectric junctions},\ }\href@noop {} {\bibfield  {journal}
  {\bibinfo  {journal} {Physical Review E}\ }\textbf {\bibinfo {volume} {99}},\
  \bibinfo {pages} {062141} (\bibinfo {year} {2019})}\BibitemShut {NoStop}%
\bibitem [{\citenamefont {Michał~Horodecki}(2013)}]{Horodecki2013}%
  \BibitemOpen
  \bibfield  {author} {\bibinfo {author} {\bibfnamefont {J.~O.}\ \bibnamefont
  {Michał~Horodecki}},\ }\bibfield  {title} {\bibinfo {title} {Fundamental
  limitations for quantum and nanoscale thermodynamics},\ }\href@noop {}
  {\bibfield  {journal} {\bibinfo  {journal} {Nature Communications}\ }\textbf
  {\bibinfo {volume} {4}},\ \bibinfo {pages} {2059} (\bibinfo {year}
  {2013})}\BibitemShut {NoStop}%
\bibitem [{\citenamefont {Segal}(2018)}]{Sega2018}%
  \BibitemOpen
  \bibfield  {author} {\bibinfo {author} {\bibfnamefont {D.}~\bibnamefont
  {Segal}},\ }\bibfield  {title} {\bibinfo {title} {Current fluctuations in
  quantum absorption refrigerators},\ }\href@noop {} {\bibfield  {journal}
  {\bibinfo  {journal} {Physical Review E}\ }\textbf {\bibinfo {volume} {97}},\
  \bibinfo {pages} {052145} (\bibinfo {year} {2018})}\BibitemShut {NoStop}%
\bibitem [{\citenamefont {Brandner}\ \emph
  {et~al.}(2018{\natexlab{b}})\citenamefont {Brandner}, \citenamefont
  {Hanazato},\ and\ \citenamefont {Saito}}]{Brandner2018}%
  \BibitemOpen
  \bibfield  {author} {\bibinfo {author} {\bibfnamefont {K.}~\bibnamefont
  {Brandner}}, \bibinfo {author} {\bibfnamefont {T.}~\bibnamefont {Hanazato}},\
  and\ \bibinfo {author} {\bibfnamefont {K.}~\bibnamefont {Saito}},\ }\bibfield
   {title} {\bibinfo {title} {Thermodynamic bounds on precision in ballistic
  multiterminal transport},\ }\href@noop {} {\bibfield  {journal} {\bibinfo
  {journal} {Phys. Rev. Lett.}\ }\textbf {\bibinfo {volume} {120}},\ \bibinfo
  {pages} {090601} (\bibinfo {year} {2018}{\natexlab{b}})}\BibitemShut
  {NoStop}%
\bibitem [{\citenamefont {Ptaszy\ifmmode~\acute{n}\else
  \'{n}\fi{}ski}(2018)}]{Ptaszy2018}%
  \BibitemOpen
  \bibfield  {author} {\bibinfo {author} {\bibfnamefont {K.}~\bibnamefont
  {Ptaszy\ifmmode~\acute{n}\else \'{n}\fi{}ski}},\ }\bibfield  {title}
  {\bibinfo {title} {Coherence-enhanced constancy of a quantum thermoelectric
  generator},\ }\href@noop {} {\bibfield  {journal} {\bibinfo  {journal} {Phys.
  Rev. B}\ }\textbf {\bibinfo {volume} {98}},\ \bibinfo {pages} {085425}
  (\bibinfo {year} {2018})}\BibitemShut {NoStop}%
\bibitem [{\citenamefont {Saryal}\ \emph {et~al.}(2019)\citenamefont {Saryal},
  \citenamefont {Friedman}, \citenamefont {Segal},\ and\ \citenamefont
  {Agarwalla}}]{Sary2019}%
  \BibitemOpen
  \bibfield  {author} {\bibinfo {author} {\bibfnamefont {S.}~\bibnamefont
  {Saryal}}, \bibinfo {author} {\bibfnamefont {H.~M.}\ \bibnamefont
  {Friedman}}, \bibinfo {author} {\bibfnamefont {D.}~\bibnamefont {Segal}},\
  and\ \bibinfo {author} {\bibfnamefont {B.~K.}\ \bibnamefont {Agarwalla}},\
  }\bibfield  {title} {\bibinfo {title} {Thermodynamic uncertainty relation in
  thermal transport},\ }\href@noop {} {\bibfield  {journal} {\bibinfo
  {journal} {Physical Review E}\ }\textbf {\bibinfo {volume} {100}},\ \bibinfo
  {pages} {042101} (\bibinfo {year} {2019})}\BibitemShut {NoStop}%
\bibitem [{\citenamefont {Buffoni}\ and\ \citenamefont
  {Campisi}(2020)}]{Buff2020}%
  \BibitemOpen
  \bibfield  {author} {\bibinfo {author} {\bibfnamefont {L.}~\bibnamefont
  {Buffoni}}\ and\ \bibinfo {author} {\bibfnamefont {M.}~\bibnamefont
  {Campisi}},\ }\bibfield  {title} {\bibinfo {title} {Thermodynamics of a
  quantum annealer},\ }\href@noop {} {\bibfield  {journal} {\bibinfo  {journal}
  {Quantum Science and Technology}\ }\textbf {\bibinfo {volume} {5}},\ \bibinfo
  {pages} {035013} (\bibinfo {year} {2020})}\BibitemShut {NoStop}%
\bibitem [{\citenamefont {Benenti}\ \emph {et~al.}(2020)\citenamefont
  {Benenti}, \citenamefont {Casati},\ and\ \citenamefont {Wang}}]{Bene2020}%
  \BibitemOpen
  \bibfield  {author} {\bibinfo {author} {\bibfnamefont {G.}~\bibnamefont
  {Benenti}}, \bibinfo {author} {\bibfnamefont {G.}~\bibnamefont {Casati}},\
  and\ \bibinfo {author} {\bibfnamefont {J.}~\bibnamefont {Wang}},\ }\bibfield
  {title} {\bibinfo {title} {Power, efficiency, and fluctuations in
  steady-state heat engines},\ }\href@noop {} {\bibfield  {journal} {\bibinfo
  {journal} {Physical Review E}\ }\textbf {\bibinfo {volume} {102}},\ \bibinfo
  {pages} {040103} (\bibinfo {year} {2020})}\BibitemShut {NoStop}%
\bibitem [{\citenamefont {Hajiloo}\ \emph {et~al.}(2020)\citenamefont
  {Hajiloo}, \citenamefont {S{\'a}nchez}, \citenamefont {Whitney},\ and\
  \citenamefont {Splettstoesser}}]{Hajiloo20}%
  \BibitemOpen
  \bibfield  {author} {\bibinfo {author} {\bibfnamefont {F.}~\bibnamefont
  {Hajiloo}}, \bibinfo {author} {\bibfnamefont {R.}~\bibnamefont
  {S{\'a}nchez}}, \bibinfo {author} {\bibfnamefont {R.~S.}\ \bibnamefont
  {Whitney}},\ and\ \bibinfo {author} {\bibfnamefont {J.}~\bibnamefont
  {Splettstoesser}},\ }\bibfield  {title} {\bibinfo {title} {Quantifying
  nonequilibrium thermodynamic operations in a multiterminal mesoscopic
  system},\ }\href@noop {} {\bibfield  {journal} {\bibinfo  {journal} {Physical
  Review B}\ }\textbf {\bibinfo {volume} {102}},\ \bibinfo {pages} {155405}
  (\bibinfo {year} {2020})}\BibitemShut {NoStop}%
\bibitem [{\citenamefont {Manzano}\ \emph
  {et~al.}(2020{\natexlab{b}})\citenamefont {Manzano}, \citenamefont
  {S{\'a}nchez}, \citenamefont {Silva}, \citenamefont {Haack}, \citenamefont
  {Brask}, \citenamefont {Brunner},\ and\ \citenamefont {Potts}}]{Manzano2020}%
  \BibitemOpen
  \bibfield  {author} {\bibinfo {author} {\bibfnamefont {G.}~\bibnamefont
  {Manzano}}, \bibinfo {author} {\bibfnamefont {R.}~\bibnamefont
  {S{\'a}nchez}}, \bibinfo {author} {\bibfnamefont {R.}~\bibnamefont {Silva}},
  \bibinfo {author} {\bibfnamefont {G.}~\bibnamefont {Haack}}, \bibinfo
  {author} {\bibfnamefont {J.~B.}\ \bibnamefont {Brask}}, \bibinfo {author}
  {\bibfnamefont {N.}~\bibnamefont {Brunner}},\ and\ \bibinfo {author}
  {\bibfnamefont {P.~P.}\ \bibnamefont {Potts}},\ }\bibfield  {title} {\bibinfo
  {title} {Hybrid thermal machines: Generalized thermodynamic resources for
  multitasking},\ }\href@noop {} {\bibfield  {journal} {\bibinfo  {journal}
  {Physical Review Research}\ }\textbf {\bibinfo {volume} {2}},\ \bibinfo
  {pages} {043302} (\bibinfo {year} {2020}{\natexlab{b}})}\BibitemShut
  {NoStop}%
\bibitem [{Note1()}]{Note1}%
  \BibitemOpen
  \bibinfo {note} {The sign convention is adopted such that a current is
  positive when it flows to the system. For instance, when a system is in the
  RP mode, it pumps heat to a hot reservoir, $J_h<0$, and it refrigerates a
  cold reservoir, $J_c>0$, by the work provided from the environment, $\protect
  \dot w<0$. In this case, the denominator of the efficiency expression is
  $-\DOTSB \sum@ \slimits@ _\alpha ^-\protect \dot {w}^{\alpha }$, and the
  numerator is $J_j\left (\protect \frac {T_{\protect \rm r}}{T_{j}}-1\right
  )$. That is, currents doing useful works are put in the numerator, while a
  current providing a resource is put in the denominator.}\BibitemShut {Stop}%
\bibitem [{\citenamefont {Dechant}(2018)}]{dechant2018multidimensional}%
  \BibitemOpen
  \bibfield  {author} {\bibinfo {author} {\bibfnamefont {A.}~\bibnamefont
  {Dechant}},\ }\bibfield  {title} {\bibinfo {title} {Multidimensional
  thermodynamic uncertainty relations},\ }\href@noop {} {\bibfield  {journal}
  {\bibinfo  {journal} {Journal of Physics A: Mathematical and Theoretical}\
  }\textbf {\bibinfo {volume} {52}},\ \bibinfo {pages} {035001} (\bibinfo
  {year} {2018})}\BibitemShut {NoStop}%
\bibitem [{Note2()}]{Note2}%
  \BibitemOpen
  \bibinfo {note} {Explicitly, \begin {eqnarray} \protect \mathbf J &= \left
  (\begin {array}{ccc} 1 & 1 & 1 \\ 0 & 1 & 0 \\ 0 & 0 & 1 \end {array} \right
  )\left ( \begin {array}{c} J_r \\ J_h \\ J_c \end {array}\right ) = \protect
  \mathcal A \protect \mathbf J_0. \end {eqnarray}}\BibitemShut {NoStop}%
\bibitem [{Note3()}]{Note3}%
  \BibitemOpen
  \bibinfo {note} {For instance, \begin {eqnarray} \protect \mathbf
  t_w^{\protect \text {EP}} =(0,0,-\protect \frac {T_r}{T_c}+1), \protect
  \,\protect \,\protect \,\protect \, \protect \mathbf t_u^{\protect \text
  {EP}} = (1,-1 + \protect \frac {T_r}{T_h}, 0), \\ \protect \mathbf
  t_w^{\protect \text {ER}} = (0, 1-\protect \frac {T_r}{T_h}, 0),\protect
  \,\protect \,\protect \,\protect \, \protect \mathbf t_u^{\protect \text
  {ER}} = (1, 0, \protect \frac {T_r}{T_c}-1). \end {eqnarray} where the ER
  mode is generating work and refrigerating, and the EP mode is generating work
  and heat pumping.}\BibitemShut {Stop}%
\bibitem [{Note4()}]{Note4}%
  \BibitemOpen
  \bibinfo {note} {For the RP operating mode of a hybrid machine, $J_u= \left [
  J_h (-1+\protect \frac {T_r}{T_h}) + J_c (\protect \frac {T_r}{T_c}-1) \right
  ]/|\protect \mathbf t_u|$ and $J_\perp = \left [-J_h (\protect \frac
  {T_r}{T_h}-1) + J_c (-1 + \protect \frac {T_r}{T_c}) \right ]/|\protect
  \mathbf t_\perp |$, where $|\protect \mathbf t_u|=|\protect \mathbf t_\perp
  |$. Hence, \begin {eqnarray} \protect \mathcal B = \left (\begin {array}{ccc}
  1 & 0 & 0 \\ 0 & \protect \qopname \relax o{cos}\alpha & \protect \qopname
  \relax o{sin}\alpha \\ 0 & -\protect \qopname \relax o{sin}\alpha & \protect
  \qopname \relax o{cos}\alpha \end {array} \right ), \end {eqnarray} where
  $\protect \qopname \relax o{cos}\alpha = (-1+\protect \frac {T_r}{T_h})
  /|\protect \mathbf t_u|$ and $\protect \qopname \relax o{sin}\alpha =
  (\protect \frac {T_r}{T_c}-1) /|\protect \mathbf t_u|$.}\BibitemShut {Stop}%
\bibitem [{Note5()}]{Note5}%
  \BibitemOpen
  \bibinfo {note} {In the RP mode, $\protect \frac {J_w}{J_u} = -(\eta
  ^{RP})^{-1} $, and $\protect \frac {J_\perp }{J_u} = -\protect \frac {1}{\eta
  ^{RP}}\left [ \eta ^P r_t - \eta ^R r_t^{-1} \right ] $, where $r_t =
  \protect \frac {T_r/T_c - 1}{-1+T_r/T_h} $. Hence, the upper bound in
  Eq.~(\ref {eq:ju}) is determined by thermal conductance tensor, temperatures
  and efficiency.}\BibitemShut {Stop}%
\bibitem [{\citenamefont {Rozhkov}\ and\ \citenamefont
  {Arovas}(1999)}]{PhysRevLett.82.2788}%
  \BibitemOpen
  \bibfield  {author} {\bibinfo {author} {\bibfnamefont {A.~V.}\ \bibnamefont
  {Rozhkov}}\ and\ \bibinfo {author} {\bibfnamefont {D.~P.}\ \bibnamefont
  {Arovas}},\ }\bibfield  {title} {\bibinfo {title} {Josephson coupling through
  a magnetic impurity},\ }\href {https://doi.org/10.1103/PhysRevLett.82.2788}
  {\bibfield  {journal} {\bibinfo  {journal} {Phys. Rev. Lett.}\ }\textbf
  {\bibinfo {volume} {82}},\ \bibinfo {pages} {2788} (\bibinfo {year}
  {1999})}\BibitemShut {NoStop}%
\bibitem [{\citenamefont {Rozhkov}\ and\ \citenamefont
  {Arovas}(2000)}]{Arovas00}%
  \BibitemOpen
  \bibfield  {author} {\bibinfo {author} {\bibfnamefont {A.}~\bibnamefont
  {Rozhkov}}\ and\ \bibinfo {author} {\bibfnamefont {D.~P.}\ \bibnamefont
  {Arovas}},\ }\bibfield  {title} {\bibinfo {title} {Interacting-impurity
  josephson junction: Variational wave functions and slave-boson mean-field
  theory},\ }\href@noop {} {\bibfield  {journal} {\bibinfo  {journal} {Physical
  Review B}\ }\textbf {\bibinfo {volume} {62}},\ \bibinfo {pages} {6687}
  (\bibinfo {year} {2000})}\BibitemShut {NoStop}%
\bibitem [{sup()}]{supp}%
  \BibitemOpen
  \href@noop {} {\bibinfo {title} {Supplemental material}},\ \bibinfo {note}
  {supplemental Material}\BibitemShut {NoStop}%
\bibitem [{\citenamefont {Katz}\ \emph {et~al.}(1983)\citenamefont {Katz},
  \citenamefont {Lebowitz},\ and\ \citenamefont {Spohn}}]{PhysRevB.28.1655}%
  \BibitemOpen
  \bibfield  {author} {\bibinfo {author} {\bibfnamefont {S.}~\bibnamefont
  {Katz}}, \bibinfo {author} {\bibfnamefont {J.~L.}\ \bibnamefont {Lebowitz}},\
  and\ \bibinfo {author} {\bibfnamefont {H.}~\bibnamefont {Spohn}},\ }\bibfield
   {title} {\bibinfo {title} {Phase transitions in stationary nonequilibrium
  states of model lattice systems},\ }\href
  {https://doi.org/10.1103/PhysRevB.28.1655} {\bibfield  {journal} {\bibinfo
  {journal} {Phys. Rev. B}\ }\textbf {\bibinfo {volume} {28}},\ \bibinfo
  {pages} {1655} (\bibinfo {year} {1983})}\BibitemShut {NoStop}%
\bibitem [{\citenamefont {Esposito}\ and\ \citenamefont
  {Schaller}(2012)}]{Esposito_2012}%
  \BibitemOpen
  \bibfield  {author} {\bibinfo {author} {\bibfnamefont {M.}~\bibnamefont
  {Esposito}}\ and\ \bibinfo {author} {\bibfnamefont {G.}~\bibnamefont
  {Schaller}},\ }\bibfield  {title} {\bibinfo {title} {Stochastic
  thermodynamics for {\textquotedblleft}maxwell demon{\textquotedblright}
  feedbacks},\ }\href {https://doi.org/10.1209/0295-5075/99/30003} {\bibfield
  {journal} {\bibinfo  {journal} {{EPL} (Europhysics Letters)}\ }\textbf
  {\bibinfo {volume} {99}},\ \bibinfo {pages} {30003} (\bibinfo {year}
  {2012})}\BibitemShut {NoStop}%
\bibitem [{\citenamefont {Rossell\'o}\ \emph {et~al.}(2017)\citenamefont
  {Rossell\'o}, \citenamefont {L\'opez},\ and\ \citenamefont
  {Platero}}]{Guillem2017}%
  \BibitemOpen
  \bibfield  {author} {\bibinfo {author} {\bibfnamefont {G.}~\bibnamefont
  {Rossell\'o}}, \bibinfo {author} {\bibfnamefont {R.}~\bibnamefont
  {L\'opez}},\ and\ \bibinfo {author} {\bibfnamefont {G.}~\bibnamefont
  {Platero}},\ }\bibfield  {title} {\bibinfo {title} {Chiral maxwell demon in a
  quantum hall system with a localized impurity},\ }\href@noop {} {\bibfield
  {journal} {\bibinfo  {journal} {Phys. Rev. B}\ }\textbf {\bibinfo {volume}
  {96}},\ \bibinfo {pages} {075305} (\bibinfo {year} {2017})}\BibitemShut
  {NoStop}%
\bibitem [{\citenamefont {Tabatabaei}\ \emph {et~al.}(2020)\citenamefont
  {Tabatabaei}, \citenamefont {S\'anchez}, \citenamefont {Yeyati},\ and\
  \citenamefont {S\'anchez}}]{Tabatabaei2020}%
  \BibitemOpen
  \bibfield  {author} {\bibinfo {author} {\bibfnamefont {S.~M.}\ \bibnamefont
  {Tabatabaei}}, \bibinfo {author} {\bibfnamefont {D.}~\bibnamefont
  {S\'anchez}}, \bibinfo {author} {\bibfnamefont {A.~L.}\ \bibnamefont
  {Yeyati}},\ and\ \bibinfo {author} {\bibfnamefont {R.}~\bibnamefont
  {S\'anchez}},\ }\bibfield  {title} {\bibinfo {title} {Andreev-coulomb drag in
  coupled quantum dots},\ }\href@noop {} {\bibfield  {journal} {\bibinfo
  {journal} {Phys. Rev. Lett.}\ }\textbf {\bibinfo {volume} {125}},\ \bibinfo
  {pages} {247701} (\bibinfo {year} {2020})}\BibitemShut {NoStop}%
\bibitem [{\citenamefont {Maes}(2021)}]{Christian2021}%
  \BibitemOpen
  \bibfield  {author} {\bibinfo {author} {\bibfnamefont {C.}~\bibnamefont
  {Maes}},\ }\bibfield  {title} {\bibinfo {title} {{Local detailed balance}},\
  }\href {https://doi.org/10.21468/SciPostPhysLectNotes.32} {\bibfield
  {journal} {\bibinfo  {journal} {SciPost Phys. Lect. Notes}\ ,\ \bibinfo
  {pages} {32}} (\bibinfo {year} {2021})}\BibitemShut {NoStop}%
\bibitem [{\citenamefont {Esposito}\ and\ \citenamefont
  {Mukamel}(2006)}]{esposito2006fluctuation}%
  \BibitemOpen
  \bibfield  {author} {\bibinfo {author} {\bibfnamefont {M.}~\bibnamefont
  {Esposito}}\ and\ \bibinfo {author} {\bibfnamefont {S.}~\bibnamefont
  {Mukamel}},\ }\bibfield  {title} {\bibinfo {title} {Fluctuation theorems for
  quantum master equations},\ }\href@noop {} {\bibfield  {journal} {\bibinfo
  {journal} {Physical Review E}\ }\textbf {\bibinfo {volume} {73}},\ \bibinfo
  {pages} {046129} (\bibinfo {year} {2006})}\BibitemShut {NoStop}%
\bibitem [{\citenamefont {Shiraishi}\ and\ \citenamefont
  {Saito}(2019)}]{shiraishi2019fundamental}%
  \BibitemOpen
  \bibfield  {author} {\bibinfo {author} {\bibfnamefont {N.}~\bibnamefont
  {Shiraishi}}\ and\ \bibinfo {author} {\bibfnamefont {K.}~\bibnamefont
  {Saito}},\ }\bibfield  {title} {\bibinfo {title} {Fundamental relation
  between entropy production and heat current},\ }\href@noop {} {\bibfield
  {journal} {\bibinfo  {journal} {Journal of Statistical Physics}\ }\textbf
  {\bibinfo {volume} {174}},\ \bibinfo {pages} {433} (\bibinfo {year}
  {2019})}\BibitemShut {NoStop}%
\bibitem [{\citenamefont {Maisi}\ \emph {et~al.}(2014)\citenamefont {Maisi},
  \citenamefont {Kambly}, \citenamefont {Flindt},\ and\ \citenamefont
  {Pekola}}]{Maisi2014}%
  \BibitemOpen
  \bibfield  {author} {\bibinfo {author} {\bibfnamefont {V.~F.}\ \bibnamefont
  {Maisi}}, \bibinfo {author} {\bibfnamefont {D.}~\bibnamefont {Kambly}},
  \bibinfo {author} {\bibfnamefont {C.}~\bibnamefont {Flindt}},\ and\ \bibinfo
  {author} {\bibfnamefont {J.~P.}\ \bibnamefont {Pekola}},\ }\bibfield  {title}
  {\bibinfo {title} {Full counting statistics of andreev tunneling},\
  }\href@noop {} {\bibfield  {journal} {\bibinfo  {journal} {Phys. Rev. Lett.}\
  }\textbf {\bibinfo {volume} {112}},\ \bibinfo {pages} {036801} (\bibinfo
  {year} {2014})}\BibitemShut {NoStop}%
\bibitem [{\citenamefont {Flindt}\ \emph {et~al.}(2008)\citenamefont {Flindt},
  \citenamefont {Novotn\'y}, \citenamefont {Braggio}, \citenamefont
  {Sassetti},\ and\ \citenamefont {Jauho}}]{Flindt2008}%
  \BibitemOpen
  \bibfield  {author} {\bibinfo {author} {\bibfnamefont {C.}~\bibnamefont
  {Flindt}}, \bibinfo {author} {\bibfnamefont {T.~c.~v.}\ \bibnamefont
  {Novotn\'y}}, \bibinfo {author} {\bibfnamefont {A.}~\bibnamefont {Braggio}},
  \bibinfo {author} {\bibfnamefont {M.}~\bibnamefont {Sassetti}},\ and\
  \bibinfo {author} {\bibfnamefont {A.-P.}\ \bibnamefont {Jauho}},\ }\bibfield
  {title} {\bibinfo {title} {Counting statistics of non-markovian quantum
  stochastic processes},\ }\href@noop {} {\bibfield  {journal} {\bibinfo
  {journal} {Phys. Rev. Lett.}\ }\textbf {\bibinfo {volume} {100}},\ \bibinfo
  {pages} {150601} (\bibinfo {year} {2008})}\BibitemShut {NoStop}%
\bibitem [{\citenamefont {Levitov}\ \emph {et~al.}(1996)\citenamefont
  {Levitov}, \citenamefont {Lee},\ and\ \citenamefont {Lesovik}}]{Levitov1996}%
  \BibitemOpen
  \bibfield  {author} {\bibinfo {author} {\bibfnamefont {L.~S.}\ \bibnamefont
  {Levitov}}, \bibinfo {author} {\bibfnamefont {H.}~\bibnamefont {Lee}},\ and\
  \bibinfo {author} {\bibfnamefont {G.~B.}\ \bibnamefont {Lesovik}},\
  }\bibfield  {title} {\bibinfo {title} {Electron counting statistics and
  coherent states of electric current},\ }\href@noop {} {\bibfield  {journal}
  {\bibinfo  {journal} {Journal of Mathematical Physics}\ }\textbf {\bibinfo
  {volume} {37}},\ \bibinfo {pages} {4845} (\bibinfo {year}
  {1996})}\BibitemShut {NoStop}%
\bibitem [{\citenamefont {Lesovik}\ and\ \citenamefont
  {Chtchelkatchev}(2003)}]{Levitov2003}%
  \BibitemOpen
  \bibfield  {author} {\bibinfo {author} {\bibfnamefont {G.}~\bibnamefont
  {Lesovik}}\ and\ \bibinfo {author} {\bibfnamefont {N.}~\bibnamefont
  {Chtchelkatchev}},\ }\bibfield  {title} {\bibinfo {title} {Quantum and
  classical binomial distributions for the charge transmitted through
  coherent},\ }\href@noop {} {\bibfield  {journal} {\bibinfo  {journal} {JEPT
  Lett.}\ }\textbf {\bibinfo {volume} {77}},\ \bibinfo {pages} {393} (\bibinfo
  {year} {2003})}\BibitemShut {NoStop}%
\bibitem [{\citenamefont {Drazin}(1958)}]{10.2307/2308576}%
  \BibitemOpen
  \bibfield  {author} {\bibinfo {author} {\bibfnamefont {M.~P.}\ \bibnamefont
  {Drazin}},\ }\bibfield  {title} {\bibinfo {title} {Pseudo-inverses in
  associative rings and semigroups},\ }\href
  {http://www.jstor.org/stable/2308576} {\bibfield  {journal} {\bibinfo
  {journal} {The American Mathematical Monthly}\ }\textbf {\bibinfo {volume}
  {65}},\ \bibinfo {pages} {506} (\bibinfo {year} {1958})}\BibitemShut
  {NoStop}%
\end{thebibliography}%

\newpage

\section{Supplemental Materials}
\hfill \break 
\tableofcontents


\section{A double quantum dot system}
\label{sec:1}
In this section we provide the detailed expressions of eigenenergies and corresponding eigenmodes of the double quantum dot system. We allow the double occupancy in the top quantum dot, and single occupancy in the bottom quantum dot. There is an onsite interaction in the top quantum dot ($U_t$), and an intra-dot interaction ($U_{tb}$) is present by a capacitive coupling. The bottom dot is coupled to a hot reservoir and cold reservoir, and their chemical potential difference ($\mu_h = V_{\rm bias}/2$ and $\mu_c = -V_{\rm bias}/2$) makes a electrical bias across the bottom quantum dot. The top quantum dot is coupled to one metallic reference lead with temperature $T_r$ and one superconducting lead. The latter provides the superconducting proximity effect. We consider the regime where the system-lead coupling is small, and they will be treated perturbatively. First, the Hamiltonian of the double quantum dot system is:

\begin{eqnarray}
\calh_{\rm DQD}^{\rm eff} &=& \epsilon_b d_b^{\dag}d_b + \sum_{\sm\in \{ \uparrow,\downarrow\}} \epsilon_{t} d_{t\sm}^{\dag}d_{t\sm} + \Gamma_S \left(d_{t\up}^{\dag} d_{t\down}^{\dag} + \rm{H.c.}\right) \nonumber
           \\ &&+ U_{tb} d_b^{\dag}d_b \sum_{\sm \in \{ \uparrow,\downarrow\}} d_{t\sm}^{\dag}d_{t\sm} + U_t d_{t\up}^{\dag}d_{t\up}d_{t\down}^{\dag}d_{t\down}.
\label{eq:heff}
\end{eqnarray}
where $\epsilon_{t,b}$ is the chemical potential of top and bottom quantum dot, respectively. $\Gamma_S$ is the coupling strength between the top quantum dot and the superconducting lead.  To diagonalize $\calh_{\rm DQD}^{\rm eff}$, we consider the basis, 
$|n_b,n_{t\up},n_{t\down}\rangle = \left(d_b^{\dag}\right)^{n_b}\left(d_{t\up}^{\dag}\right)^{n_{t\up}}\left(d_{t\down}^{\dag}\right)^{n_{t\down}}|0\rangle$,
where $n_b=0,1$ and $n_{t\sm}=0,1$. Thus, there are total eight states in the double quantum dot system. Specifically,

\begin{figure*}[t]
\centering
\includegraphics[width=.32\linewidth]{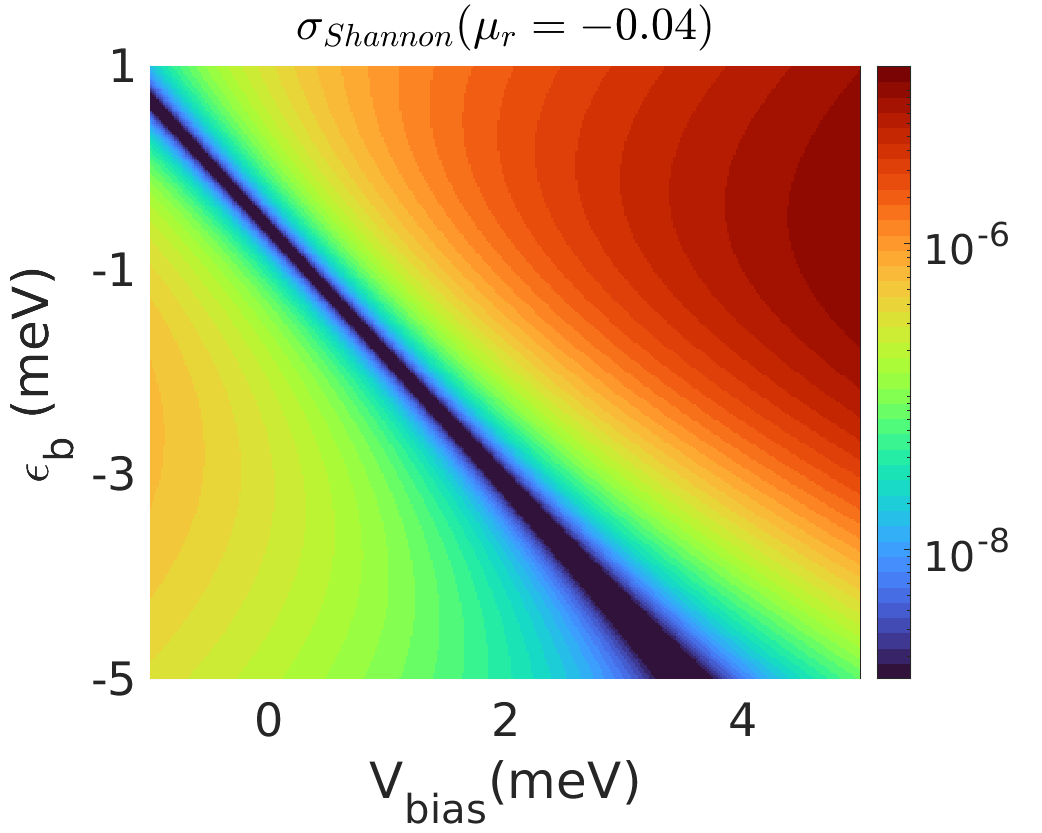}
\includegraphics[width=.32\linewidth]{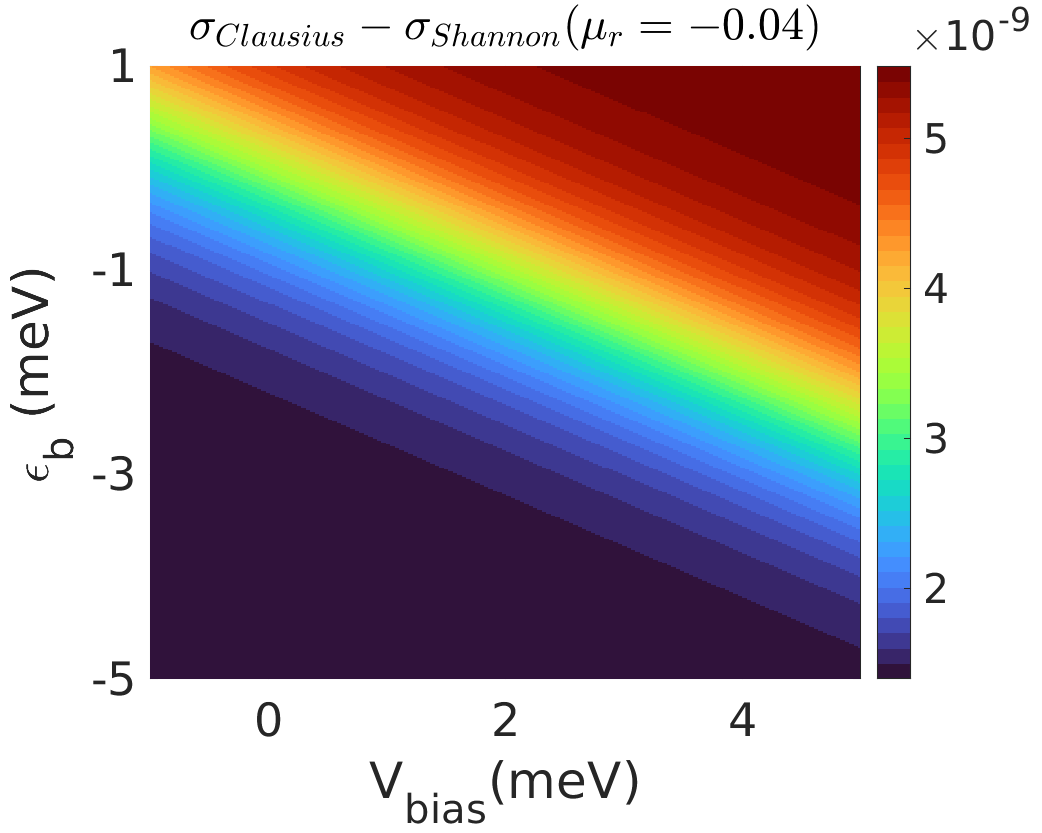}
\includegraphics[width=.32\linewidth]{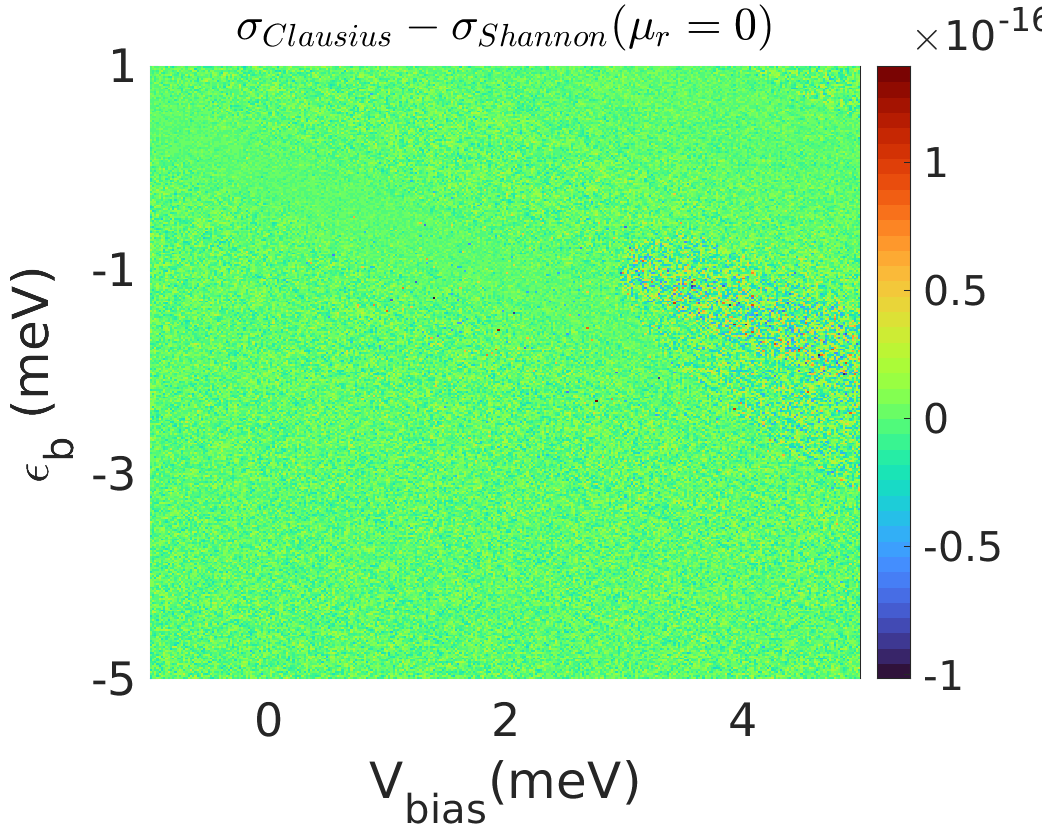}
\caption{The entropy production rate according to the Shannon expression Eq.(\ref{eq:shann}) (left panel), $\sigma_S$. The difference between two expressions of entropy production rate, $\sigma_C-\sigma_S$, is plotted (center panel). When the local detailed balance is present, $\mu_r=0$, the difference $\sigma_C-\sigma_S$ is plotted (right panel), showing that $\sigma_S= \sigma_C $ within a numerical error. 
}
\label{fig:S_C}
\end{figure*}

\begin{align}
&|0,\up\rangle = |0,1,0\rangle, \quad \text{with} \quad E_{\ket{0,\up}} = \veps_t 
\\
&|0,\down\rangle = |0,0,1\rangle, \quad \text{with} \quad E_{\ket{0,\down}} = \veps_t 
\\
&|1,\up\rangle = |1,1,0\rangle, \quad \text{with} \quad E_{\ket{1,\up}} = \veps_b + \veps_t + U_{tb}
\\
&|1,\down\rangle = |1,0,1\rangle, \quad \text{with} \quad E_{\ket{1,\down}} = \veps_b + \veps_t + U_{tb}
\end{align}
\begin{align}\label{eq:eig}
|0,+\rangle &= u_{0,+}|0,0,0\rangle + v_{0,+}|0,1,1\rangle,  \nonumber\\
&\quad \quad\quad \text{with} \quad E_{\ket{0,+}} = \veps_0 - \sqrt{\veps_0^2 + \Gamma_S^2}, \\
|0,-\rangle &= u_{0,-}|0,0,0\rangle + v_{0,-}|0,1,1\rangle,  \nonumber\\
&\quad \quad\quad \text{with} \quad E_{\ket{0,-} }= \veps_0 + \sqrt{\veps_0^2 + \Gamma_S^2},
\\
|1,+\rangle &= u_{1,+}|1,0,0\rangle + v_{1,+}|1,1,1\rangle,  \nonumber\\
&\quad\quad \quad \text{with} \quad E_{\ket{1,+}} = \veps_b + \veps_1 - \sqrt{\veps_1^2 + \Gamma_S^2},
\\
|1,-\rangle &= u_{1,-}|1,0,0\rangle + v_{1,-}|1,1,1\rangle ,  \nonumber\\
& \quad\quad\quad \text{with} \quad E_{\ket{1,-} }= \veps_b + \veps_1 + \sqrt{\veps_1^2 + \Gamma_S^2}, \label{eq:eig9}
\end{align}
where we introduce the notation used in the main text. The coefficients of the superconducting states are: 
\beq
u_{n_b,\pm} = \frac{\veps_{n_b} \pm \sqrt{\veps_{n_b}^2 + \Gamma_S^2}}{\caln_{n_b,\pm}},
\quad
v_{n_b,\pm} = -\frac{\Gamma_S}{\caln_{n_b,\pm}},
\edq
with  $u_{n_b,\pm}^2 + v_{n_b,\pm}^2 = 1$. And,
\beq
\veps_{n_b} = \veps_t + U_p/2 + n_b U_{tb} \quad \text{for $n_b=0,1$}
\quad
\edq
The normalization factor is given by
\beq
\caln_{n_b,\pm} = \sqrt{\left(\veps_{n_b} \pm \sqrt{\veps_{n_b}^2 + \Gamma_S^2}\right)^2 + \Gamma_S^2}
\edq

For later use, we specify the energy difference between states by the transition made by a coupled lead when $\Gamma_S \ll U_t, U_{tb}, \epsilon_{t}$, which is the working regime of our study: $\Gamma_S=0.2$, $U_t=1$, $U_{tb}=2$, $\epsilon_t=4$, and $\epsilon_b \in [-5,1]$ in the main text. The transitions between the eight states are induced by the leads (reference, hot, and cold reservoirs). When the transition involves an energy difference, the leads provide or absorb the extra energy. In particular, the transitions induced by  hot or cold lead are the following (they make change in the particle number sitting in  the bottom quantum dot, $n_b$): 
\begin{eqnarray} \label{eq:13}
&E_{1,-} - E_{0,+} &\simeq \epsilon_b+2\epsilon_t +U_t+2U_{tb}, \\
&E_{1,-} - E_{0,-} &\simeq \epsilon_b + 2U_{tb}, \\
&E_{1,\uparrow} - E_{0,\uparrow} &\simeq \epsilon_b +U_{tb}, \\
&E_{1,\downarrow} - E_{0,\downarrow} &\simeq \epsilon_b +U_{tb}, \\
&E_{1,+} - E_{0,+} &\simeq \epsilon_b, \\
&E_{1,+} - E_{0,-} &\simeq \epsilon_b-2\epsilon_t-U_t, \label{eq:18}
\end{eqnarray}
where the energy difference is written in such a way that 
from the top to bottom the energy difference decreases. They all share the energy $\epsilon_b$, in addition to the onsite and inter-dot interaction energy $U_{t,tb}$. 
Later, we introduce the energy dependent system-(hot,cold) lead coupling strength to operate the hybrid thermal machine, see Sec.\ref{sec:hybrid}.

\section{Entropy production: Shannon and Clausius}
\label{sec:entropy}
The entropy is the number of accessible states in reservoirs, which is in our case the three leads. When heat flows into a reservoir, the particles in the reservoir can access more number of states and as a result the entropy increases. The entropy production therefore can be expressed in terms of heat currents, and this is the Clausius entropy $\sigma_{C}$: 
\begin{align}
    \sigma_C = \sum_{l=r,h,c} -\frac{J_l}{T_l},
\end{align}
where the (-) sign is because we adopt the sign convention of current in such way that it is positive when it flows to the system. 

On the other and, the Shannon entropy production is obtained from the transition rates between states of system, which are specified in Sec.\ref{sec:1}, and the probability density~\cite{esposito2006fluctuation}: 
\begin{align} \label{eq:shann}
    \sigma_S = - \sum_{m,n} \gamma_{mn} \rho_n \ln \left( \frac{\gamma_{mn} \rho_n}{\gamma_{nm} \rho_m} \right).
\end{align}
The two expressions of the entropy production in identical in the presence of the local detailed balance (for example, see Sec.2.2.4 of Ref.~\cite{shiraishi2019fundamental}). In our double quantum dot system, the local detailed balance can be broken by introducing nonzero chemical potential $\mu_r$ at the reference frame, which is coupled to the top quantum dot (See Sec.\ref{sec:ldb}). In such a case the Clausius ($\sigma_C\equiv\sigma_{Clausius}$) and Shannon entropies ($\sigma_S\equiv\sigma_{Shannon}$) differ as shown in Fig. \ref{fig:S_C} for our setup.

\begin{figure*}[t]
\centering
\includegraphics[width=.34\linewidth]{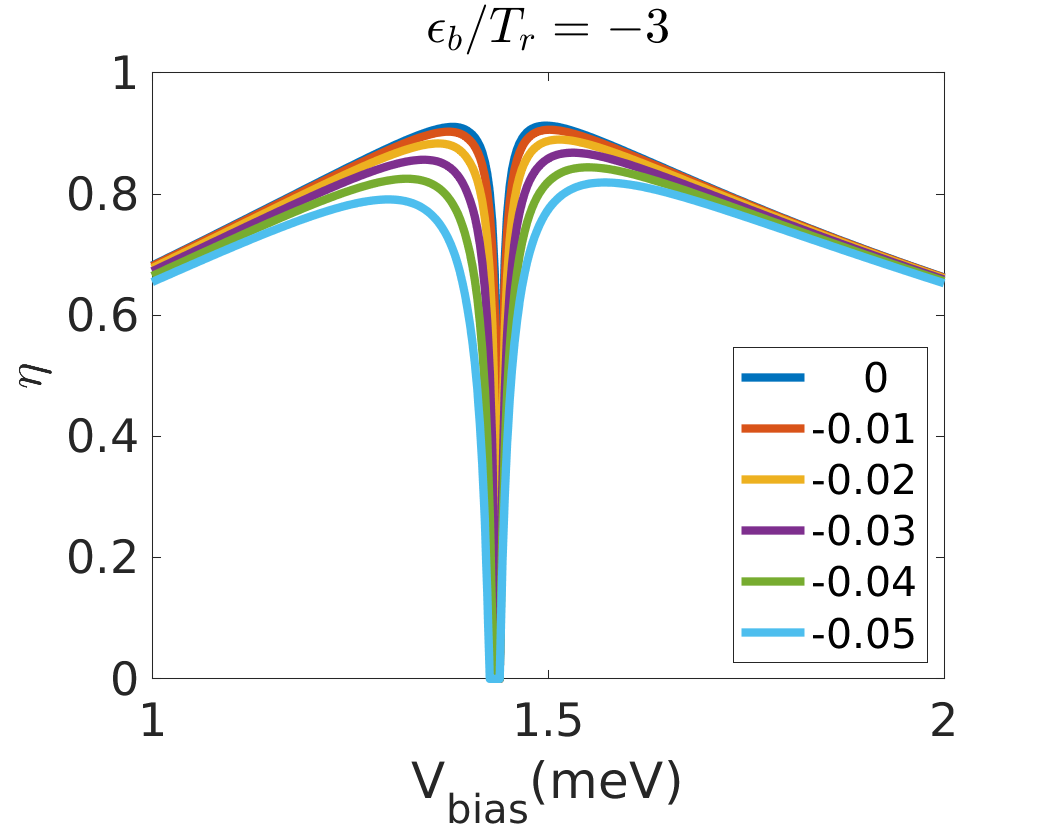}
\includegraphics[width=.34\linewidth]{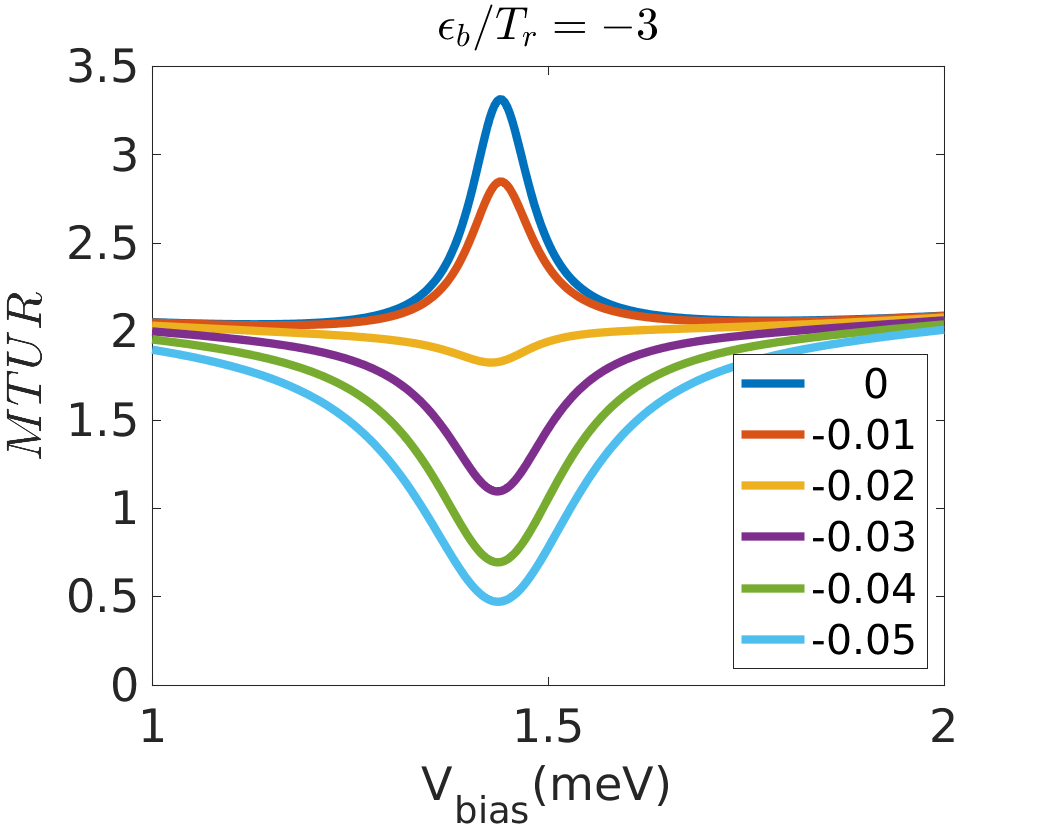}
\caption{Thermal efficiency (left) and MTUR value $\sigma \Delta P$ (right) are plotted at chemical potential of the reference reservoir $\mu_r/T_r = 0,-0.01,-0.02,-0.03$ (indicated in the legend).}
\label{fig:eta_tur}
\end{figure*}

\section{The local detailed balance}
\label{sec:ldb}
The local detailed balance (LDB) is broken in the transition which involves the change of particle number in the top quantum dot when $\mu_r \neq 0$. The transition is induced by the reference lead. Below, we consider the transition from $\ket{n_b=0, \sigma \in \{ \uparrow, \downarrow\}}$ to $\ket{n_b=0, +}$. The similar argument follows for the transition $\ket{n_b=1, \sigma \in \{ \uparrow, \downarrow\}}$ to $\ket{n_b=1, \pm}$. The forward transition rate is (the transition rate below is shown in Eq.~(\ref{eq:master1})):
\begin{align}
 &\gamma^r_{\ket{0,+} \leftarrow \ket{0,\sigma}} \nonumber \\
&= 
\gamma^{(e)} _{\ket{0,+} \leftarrow \ket{0,\sigma}} + \gamma^{(h)}_{\ket{0,+} \leftarrow \ket{0,\sigma}}, \nonumber \\ 
&= \Gamma_r \left( |v_{0+}|^2 f^e(\Delta E) + |u_{0+}|^2 f^h(-\Delta E) \right), \nonumber \\
&= \Gamma_r \left[ \frac{|v_{0+}|^2}{e^{(\Delta E-\mu_r)/T_r}+1} + \frac{ |u_{0+}|^2}{e^{(\Delta E + \mu_r)/T_r}+1}\right],
\end{align}
where $\Delta E = E_{\ket{0,+}} - E_{\ket{0,\sigma}}$,  $ f^e(\Delta E) = (e^{(\Delta E - \mu_r)/T_r}+1)^{-1} $, and $f^h(-\Delta E) =  1- (e^{(-\Delta E - \mu_r)/T_r}+1)^{-1}$. 
(Energies are written in unit of temperature of reference reservoir, $T_r$). There are two scattering processes: one is to add electron to the top quantum dot that comes with probability $|v_{0+}|^2$ and distribution $f^e(\Delta E)$, and the second term is to remove an electron from the top quantum dot that comes with probability $|u_{0+}|^2$ and distribution $f^h(-\Delta E) = 1 - f^e(-\Delta E)$. For the second process, an electron is going into the reference lead with the energy $\Delta E$ below $\mu_r$.
The forward transition rate is similarly written (the transition rate below is shown in Eq.~(\ref{qme_d})):
\begin{align}
& \gamma^r_{ \ket{0,\sigma} \leftarrow\ket{0,+}} \nonumber \\
&= \gamma^{(e)} _{ \ket{0,\sigma} \leftarrow\ket{0,+}} + \gamma^{(h)} _{ \ket{0,\sigma} \leftarrow\ket{0,+}}, \nonumber  \\ 
& = \Gamma_r \left( |v_{0+}|^2 f^e(-\Delta E) + |u_{0+}|^2 f^h(\Delta E) \right), \nonumber \\
&= \Gamma_r \left[ \frac{|v_{0+}|^2}{e^{(-\Delta E-\mu_r)/T_r}+1} + \frac{ |u_{0+}|^2}{e^{(-\Delta E + \mu_r)/T_r}+1}\right],
\end{align}

The ratio between the two transition rate is generally dependent of $|v_{0+}|^2$ and $|u_{0+}|^2$ for $\mu_r \neq 0$, therefore the local detailed balance is broken:  
\begin{align} \label{eq:ldb_sc}
\frac{\gamma_B}{\gamma_F} &= \frac{\gamma^r_{ \ket{0,\sigma} \leftarrow\ket{0,+}}}{\gamma^r_{\ket{0,+} \leftarrow \ket{0,\sigma}}} , \nonumber \\
 &= \frac{|v_{0+}|^2 f^e(-\Delta E) + |u_{0+}|^2 f^h(\Delta E)}{|v_{0+}|^2 f^e(\Delta E) + |u_{0+}|^2 f^h(-\Delta E) }, \nonumber \\
 &=e^{\frac{\Delta E}{T_r}}\left[ \frac{\cosh\theta_\gamma +e^{-\frac{\Delta E}{T_r}}\cosh (\theta_\gamma-\theta_{\mu_r})}{\cosh(\theta_\gamma+\theta_{\mu_r}) +e^{-\frac{\Delta E}{T_r}}\cosh \theta_\gamma} \right],
\end{align}
where $\cosh \theta_\gamma = \frac{1}{2}(\left|\frac{v_{n_b,\pm}}{u_{n_b,\pm}} \right|^2+ \left|\frac{u_{n_b,\pm}}{v_{n_b,\pm}} \right|^2)$ and $\cosh\theta_{\mu_r}=\frac{1}{2}(e^{\frac{\mu_r}{T_r}} + e^{-\frac{\mu_r}{T_r}} )$.  If $\mu_r = 0$, the local detailed balance is preserved. Since $\theta_{\mu_r}=0$, 
\begin{align}
\frac{\gamma_B}{\gamma_F} = \left.\frac{\gamma_{ \ket{0,\sigma} \leftarrow\ket{0,+}}}{\gamma_{{\ket{0,+} \leftarrow \ket{0,\sigma}}}} \right|_{\mu_r=0} = e^{\Delta E /T_r}. 
\end{align}
This implies that for $\Delta E >0$ the transition rate of the forward transition is exponentially small in $\Delta E$ than that of the backward transition. 
Due to the breakdown of the local detailed balance when $\mu_r\neq 0$ the multidimensional thermodynamic uncertainty relation (MTUR) is not satisfied as shown in Fig. \ref{fig:eta_tur} where the efficiency (left figure) and the MTUR (right figure) are shown. For $\mu_r\in [0.01,0.05]$ the MTUR are violated whereas the efficiency remains almost untouchable (see left figure in Fig. \ref{fig:eta_tur}).

Note that the role of superconducting lead is crucial for the breakdown of LDB. Due to the proximity effect, the two states in Eq.(\ref{eq:eig}-\ref{eq:eig9}) with even number of  particle difference are coupled. A situation without superconductivity can be simulated by setting either $v_{0+}=0$ or $u_{0+}=0$. In this case, $\theta_{\gamma} \rightarrow \infty$ in Eq.(\ref{eq:ldb_sc}) and the LDB is recovered regardless of the chemical potential in the reference lead.

\begin{figure*}[t]
\centering
\includegraphics[width=.9\linewidth]{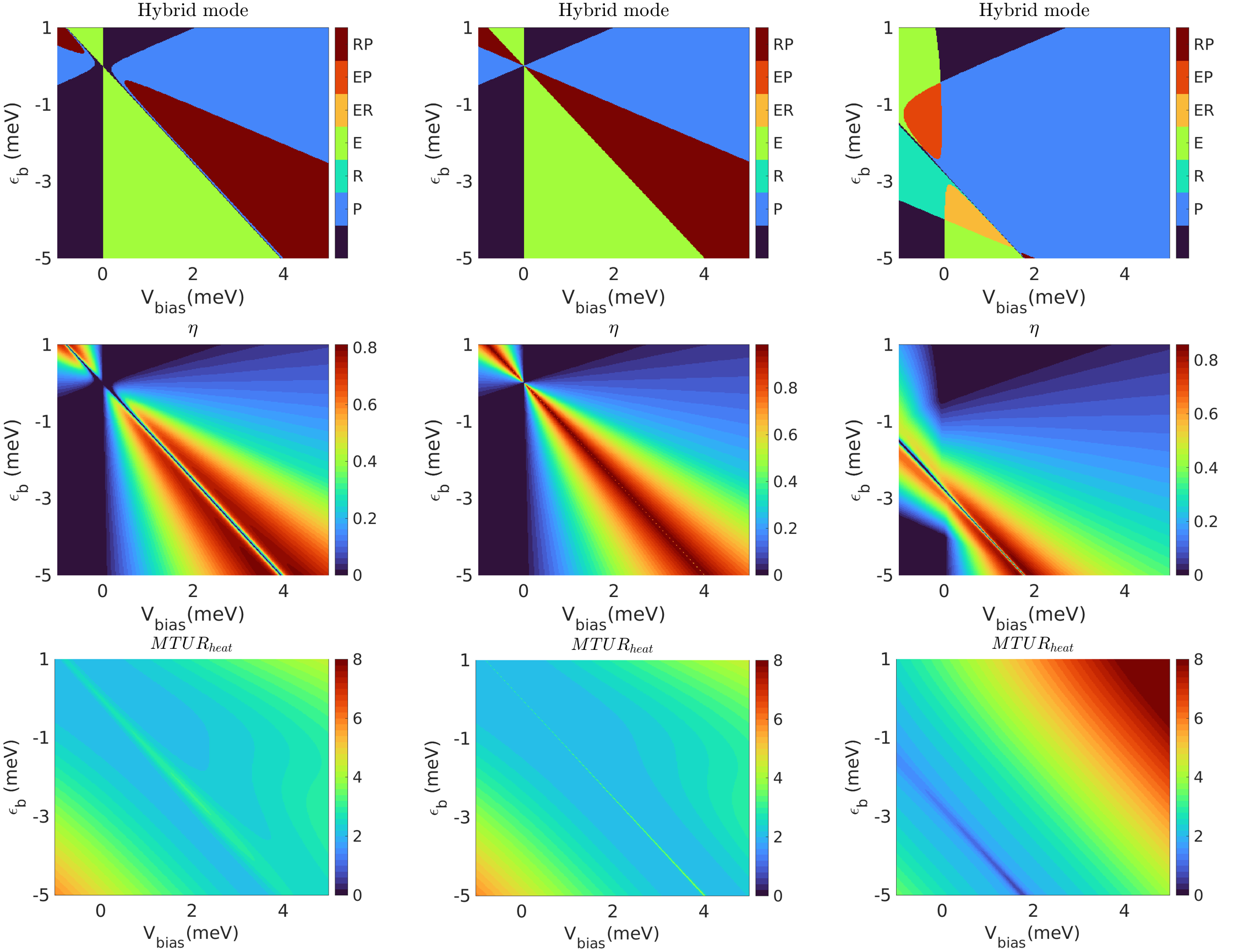}
\caption{Hybrid machine mode (first row), thermal efficiency (second row), and $\sigma \Delta P$ (third row) for different configuration of of energy-dependent lead-system coupling. (Left column): $\gamma_{0,1,2}=\Gamma_r$ corresponding to the energy independent case. (Center column): $\gamma_{0,2}=\Gamma_r$ and $\gamma_1=0$. (Right column): $\gamma_{0}=\Gamma_r$ and $\gamma_{1,2}=0$. }
\label{fig:sets}
\end{figure*}

\section{Hybrid thermal machine}
\label{sec:hybrid}

The double quantum dot system is coupled to the four lead (reference, superconducting, hot, and cold reservoir). Only three of them (except the superconducting lead) can exchange particle and energy current with the system because the superconducting gap is assumed to be infinite. Thus, our system is effectively considered as a hybrid thermal machine that is capable of one or two among three different machine modes, engine, heat pumping, and refrigerating. When the system-lead coupling strength is set energy-independent value, $\Gamma_{r,h,c}=0.05$, we find the working mode,  thermal efficiency, and $\sigma\Delta P$ as shown in the left column of Fig.~\ref{fig:sets}. The hybrid machine modes ER and EP are not present. To access the modes, it turns out~\cite{Manzano2020} that it is necessary to introduce energy-dependent system-lead coupling for hot and cold reservoir. 
\begin{align}
    &\Gamma_h (\Delta \tilde E\geq 3) = \gamma_2, \quad &\Gamma_c (\Delta \tilde E\geq 3) = \gamma_0, \\
    &\Gamma_h (3>\Delta \tilde E\geq 1) = \gamma_1, \quad &\Gamma_c (3>\Delta \tilde E\geq 1) = \gamma_1, \\
    &\Gamma_h (1>\Delta \tilde E) = \gamma_0, \quad &\Gamma_c (1>\Delta \tilde E) = \gamma_2,
\end{align}
where $ {\Delta\tilde E} = E_{\ket{f}} - E_{\ket{i}} - \epsilon_b$ (see Eq.(\ref{eq:13}-\ref{eq:18}) for the relevant transitions). For instance, we obtain the results shown in the main text  by setting $\gamma_2=\Gamma_r, \gamma_{0,1}=0$. It implies that the hot lead only exchange particles with energy $\Delta \tilde E\geq 3$, while the cold lead only exchange particles with energy $\Delta \tilde E <0$. Hence, when there is a particle flows from the hot to cold lead, the energy current is correlated with it. In the center column of Fig.~\ref{fig:sets}, we set $\gamma_{0,2}=\Gamma_r$ and $\gamma_1=0$. In the right column, we set $\gamma_{0}=\Gamma_r$ and $\gamma_{1,2}=0$, which is the opposite situation from the condition in the main text: energetic particles $\Delta \tilde E >2$ from the cold lead can tunnel to the system, while only particles with $\Delta \tilde E<0$ from the hot lead can tunnel to the system.

\section{Full Counting Statistics}

   The FCS technique is considered to derive the second cumulant for the heat and charge transport and its correlations. For such purpose we follow  Ref.~\cite{Maisi2014, Flindt2008}. We firstly solve the set of master equations and obtain the cumulant generating function \cite{Levitov1996,Levitov2003}. We first, introduce the eigenvectors $a_{\beta}$ and eigenvalues $\lambda_{\beta}$ of the matrix $\call$, defined by $\call a_{\beta} = \lambda_{\beta} a_{\beta}$. Formally we can write the solution of the master equation as $|\rho(t)\rangle = \sum_{\beta} c_{\beta} \exp\left(-\lambda_{\beta}t\right) a_{\beta}$ where the coefficients $c_{\beta}$ are determined by the initial condition. The eigenvalues $\lambda_{\beta}$ are  real numbers since $\call$ is symmetric and positive to have positive probabilities.  Besides, one eigenvalue, say $\beta=0$, must be equal to zero, $\lambda_0=0$, which corresponds to the stationary case. The stationary probabilities can be obtained from the left $\langle 0_L|$ and right $|0_R\rangle$ null eigenvectors of $\call$
corresponding to zero eigenvalue $|\rho^{(st)}\rangle = \frac{1}{\langle 0_L | 0_R\rangle}|0_R\rangle = |0_R\rangle$.  

Inasmuch as we are interested in the charge and heat noise for both  dots  we introduce the particle and energy transport on the same footing, we need to introduce both the particle-number and energy resolved density matrix
\beq
\rho_{\gamma\eta}(\{\bold N,\bold E\};t) \equiv \rho_{\gamma\eta}(\{N_r, N_h, N_c, E_r,E_h,E_c\};t)
\edq
which gives the probability of having a particle and an energy $N_\alpha$, $E_{\alpha}$ in each corresponding lead $\alpha$ by the time $t$, where the subscripts $\gamma$ and $\eta$ denote the different dot system states. In our study we focus on the dynamics of the population, thus $\left. \rho_{\gamma\eta}\right|_{\gamma=\eta}\equiv \rho_{\eta}$.  $N_r$ ($E_r$) distinguishes the charge (energy) at the reference reservoir for the top quantum dot from the charge $N_{h,c}$ ($E_{h,c}$) at the hot and cold reservoirs for the bottom quantum dot. Its Fourier transform is defined as
\beq
\tilde \rho_{\eta}(\{\chi,\kappa\};t) 
= \sum_{\{\bold N,\bold E\}} \rho_{\eta}(\{\bold N,\bold E\};t) e^{i\chi \cdot \bold N}e^{i\kappa \cdot \bold E}
\edq
where $\bold N=(N_r, N_h, N_c)$ and $\bold E = (E_r,E_h,E_c) $. The counting fields are the conjugate particle and energy variables: for the particle  $\chi=(\chi_r,\chi_h,\chi_c)$, and for the energy $\bold \kappa=(\kappa_r$, $\kappa_h,  \kappa_c)$ counting fields. After Fourier transforming $\rho_{\eta}(\{\bold N,\bold E\})$, the master equation reads
\beq
\frac{d}{dt} \tilde \rho\left(\{\chi,\kappa\};t\right) = -\call\left(\{\chi,\kappa\}\right) \tilde \rho\left(\{\chi,\kappa\};t\right)
\edq
The explicit form for $\call\left(\{\chi,\kappa\}\right)$ is written in Sec.\ref{sec:master})
Here, the particle-energy number resolved density matrix fulfils the master equation: $\frac{d}{dt} \tilde \rho(\{\chi,\kappa\};t) = -\call(\{\chi,\kappa\})\tilde \rho(\{\chi,\kappa\};t)$. 
From these equations the Liouvillian operator can be written as $\call(\{\chi, \kappa\}) = \call(\{\bold 0, \bold 0 \}) + \tilde{\call}(\{\chi, \kappa\})$. For the cumulant calculation we first compute the  projection and pseudoinverse operators $\calp = 1 - |0_R\rangle \langle 0_L|, \quad \calq = 1 - \calp, \quad \calr = \calq \call^{-1} \calq$ where the Drazine inverse \cite{10.2307/2308576} of $\call$ is denoted as $\call^{-1} $. 
For the top dot the charge and energy current at the reference lead 
it can be derived from the first order cumulants. For instance, if one is interested in computing charge current via lead $\alpha \in \{r,h,c\}$,
\begin{eqnarray}\label{Currs_N}
&&j^q_\alpha=e C^q_\alpha = -e\langle 0_L | \left. \frac{\partial \tilde{\call} (\{ \chi,\kappa\})}{\partial (i\chi_\alpha)} \right|_{\chi=\kappa=\bold 0}|0_R\rangle,
\end{eqnarray}
which simply computes the change of particle number in lead $\alpha$. 
If one is interested in computing the energy current via lead $\beta \in \{r,h,c\}$, 
\begin{eqnarray}\label{Currs_E}
&&j^e_\beta= C^e_\beta = -\langle 0_L | \left. \frac{\partial \tilde{\call} (\{ \chi,\kappa\}) }{\partial (i\kappa_\beta)}\right|_{\chi=\kappa=\bold 0}|0_R\rangle,
\end{eqnarray}
which computes the change of total energy in lead $\beta$. 
The heat current via lead $\gamma \in \{r,h,c \}$ is
\begin{align}
    j^h_\gamma &= j^e_\gamma - \mu_\alpha j^q_\gamma, \\
    &= C^e_\gamma - e\mu_\gamma C^q_\gamma, 
\end{align}

Next, let us compute the  charge and energy noise associated with lead $\alpha$ and $\beta$, respectively. For the evaluation of multidimensional TUR, one must include the cross current-current correlation among three leads, $\alpha \neq \beta$. First, the correlation of charge current via lead $\alpha$ and $\beta$ is
\begin{align}\label{mixednoise}
C^{qq}_{\alpha \beta} &= -\langle 0_L|\left[\partial_{i\chi_\alpha}\partial_{i\chi_\beta}\tilde{\call} \right]_{\bold 0}|0_R\rangle \nonumber \\
&+  \left[ \langle 0_L|\left[\partial_{i\chi_\alpha}\tilde{\call}\right]_{\bold 0}\calr\left[ \partial_{i\chi_\beta}\tilde{\call}\right]_{\bold 0}|0_R\rangle + \left(\chi_\alpha \Leftrightarrow \chi_\beta \right)  \right] \nonumber \\
\end{align}
where $\left[ \cdot \right]_{\bold 0} =\left[ \cdot \right]_{\chi=\kappa=\bold 0} $. When $\alpha=\beta$, Eq.(\ref{mixednoise}) provides the charge noise in lead $\alpha$. The correlation of energy current via lead $\alpha$ and $\beta$ is similarly written: 
\begin{align}\label{mixednoise_E}
C^{ee}_{\alpha \beta} &= -\langle 0_L|\left[\partial_{i\kappa_\alpha}\partial_{i\kappa_\beta}\tilde{\call} \right]_{\bold 0}|0_R\rangle \nonumber \\
&+  \left[ \langle 0_L|\left[\partial_{i\kappa_\alpha}\tilde{\call}\right]_{\bold 0}\calr\left[ \partial_{i\kappa_\beta}\tilde{\call}\right]_{\bold 0}|0_R\rangle + \left(\kappa_\alpha \Leftrightarrow \kappa_\beta \right)  \right], \nonumber \\
\end{align}
where the counting field is simply replaced by $\kappa_{\alpha,\beta}$. The correlation between the charge current via lead $\alpha$ and the energy current via lead $\beta$ is as follows:
\begin{align}\label{mixednoise_PE}
C^{qe}_{\alpha \beta} &= -\langle 0_L|\left[\partial_{i\chi_\alpha}\partial_{i\kappa_\beta}\tilde{\call} \right]_{\bold 0}|0_R\rangle \nonumber \\
&+  \left[ \langle 0_L|\left[\partial_{i\chi_\alpha}\tilde{\call}\right]_{\bold 0}\calr\left[ \partial_{i\kappa_\beta}\tilde{\call}\right]_{\bold 0}|0_R\rangle + \left(\chi_\alpha \Leftrightarrow \kappa_\beta \right)  \right]. \nonumber \\
\end{align}
The evaluation of heat noise can be obtained from the charge and energy cross correlations. 
\begin{align}\label{mixednoise_HH}
C^{hh}_{\alpha \beta} &= C^{ee}_{\alpha\beta} -e \mu_\beta C^{eq}_{\alpha\beta} -e \mu_\alpha C^{qe}_{\alpha\beta} +e^2\mu_\alpha \mu_\beta C^{qq}_{\alpha\beta}, 
\end{align}
which follows from the definition of heat current. The  expression above is enough for our purpose of computing the MTUR from heat currents via three leads. Note that  other combinations include
\begin{align}
C^{hq}_{\alpha \beta}  &= C^{eq}_{\alpha \beta}      - e\mu_\alpha C^{qq}_{\alpha \beta}, \\
C^{qh}_{\alpha \beta}  &= C^{qe}_{\alpha \beta}      - e\mu_\beta C^{qq}_{\alpha \beta}.  
\end{align}

\section{Quantum master equation} \label{sec:master}
In this section we provide the derivation of the master equation employed in the main text from the Lindblad equation. In the first subsection, We consider the top quantum dot coupled to one metallic lead (reference reservoir) and one superconducting lead. The master equation is derived for $\Gamma_r \ll \Gamma_S$ so that the top quantum dot remains in the superconducting phase. The dimension of Hilbert space is four, $\ket{n_{t\uparrow}} \in \{0,1\},n_{t\downarrow} \in \{0,1\} $. 

In the second subsection, the bottom quantum dot is introduced, $n_b\in \{0,1\}$, by which the dimension of the Hilbert space is doubled. There is one space with $n_b=0$, and the other space with $n_b=1$. The two spaces are coupled by hot and cold lead which exchange electron and hole with the bottom quantum dot. We work in the regime where $T_{r,h,c} \gg  \Gamma_{r,h,c}$, which allows us to neglect the off-diagonal elements of the density matrix. Note that the master equation (rate equation approach) is employed in Ref.~\cite{Tabatabaei2020}, and here we provide its justification.

\subsection{A single quantum dot}

In this section we obtain the quantum master equation for the simple case of a single quantum dot attached to two contacts, one is a metallic reservoir and the other is superconduting (N-QD-S). For such purpose we employ the quantum Lindblard formalism \cite{}. Our starting point is the effective Hamiltonian by the superconducting proximity effect  \cite{Arovas00} for the N-QD-S system valid when the largest energy scale $\Delta \rightarrow \infty$ as discussed in the main text. Such Hamiltonian is
\begin{equation}
\hat H_{n_b=0} = \epsilon_t \sum_{\sigma = \uparrow, \downarrow} d^\dagger_\sigma d_\sigma + U_t \hat n_\uparrow \hat n_\downarrow  + \Gamma_S d^\dagger _\uparrow d^\dagger _\downarrow + \Gamma_S d_\downarrow d_\uparrow, 
\end{equation}
where the subscript $n_b=0$ is used because the Hamiltonian corresponds to the situation where the particle number of the bottom quantum dot is fixed to $n_b=0$. 
Writing this in matrix form in the electron number basis, $|n_\downarrow, n_\uparrow\rangle=\{|0,0\rangle, |1,1\rangle, |1,0\rangle, |0,1\rangle \}$,

\begin{eqnarray}\label{Hnb0}
 \hat H_{n_b=0} = 
\left(\begin{array}{cccc}
0 & \Gamma_S & 0 & 0 \\
\Gamma_S & 2\epsilon_t + U_t & 0 & 0 \\
0 & 0 &\epsilon_t & 0 \\ 
0 & 0 & 0 &\epsilon_t 
\end{array}\right)
.
\end{eqnarray}
The eigenvalues of the first 2$\times$2 block are
\begin{equation}
\epsilon_\pm = \epsilon_t + \frac{U_t}{2}  \pm \sqrt{\Gamma_S^2 + \left(\epsilon_t + \frac{U_t}{2}\right)^2}.
\end{equation}
Corresponding eigenvectors are
\begin{equation}
\ket{\pm} = u_\pm \ket{0,0} + v_\pm \ket{1,1},    
\end{equation}
where 
\begin{equation}
u_\pm = \epsilon_{\pm}/\sqrt{\Gamma_S^2 + \epsilon_{\pm}^2},\quad v_{\pm} = -\Gamma_S/\sqrt{\Gamma_S^2 + \epsilon_{\pm}^2},
\end{equation}
where note that we used different sign of $v_\pm$ from the main text to simplify the notation in the following discussion. The master equation Eq.(\ref{eq:master1}) that we obtain later is independent of this choice.  
Introducing the unitary matrix
\begin{eqnarray}\nonumber
U_{0} =\left( \begin{array}{cccc}
u_+ & u_- & 0 & 0\\
v_+ & v_- & 0 & 0\\
0 & 0 & 1 & 0\\
0 & 0 & 0 & 1 
\end{array}\right), \,\,
U_{0}^\dagger = \left( \begin{array}{cccc}
u^*_+ & v^*_+ & 0 & 0 \\
u^*_- & v^*_- & 0 & 0 \\
0 & 0 & 1 & 0 \\
0 & 0 & 0 & 1 
\end{array}\right),
\end{eqnarray}
the Hamiltonian can be diagonalized:
$ D = U_{0}^\dagger H_{n_b=0} U_{0}= \text{diag} (\epsilon_+,\epsilon_-,\epsilon_t,\epsilon_t )$. 
Let us explicitly write the operators in the basis of $\ket {n_\downarrow, n_\uparrow}$: 
\begin{eqnarray}
\hat d_\uparrow =\left( \begin{array}{cccc} 
 0 & 0 & 0 & 1 \\
 0 & 0 & 0 & 0\\
 0 & 1 & 0 & 0\\
 0 & 0 & 0 & 0
\end{array}\right), \,\,\,\,
\hat d^\dagger_\uparrow = \left(\begin{array}{cccc} 
0 & 0 & 0 & 0  \\
0 & 0 & 1 & 0\\
0 & 0 & 0 & 0\\
1 & 0 & 0 & 0
\end{array}\right), \\
\hat d_\downarrow = \left( \begin{array}{cccc}
0 & 0 & 1 & 0  \\
0 & 0 & 0 & 0\\
0 & 0 & 0 & 0 \\
0 & 1 & 0 & 0
\end{array}\right), \,\,\,\,
\hat d^\dagger_\downarrow =\left(  \begin{array}{cccc}
0 & 0 & 0 & 0 \\
0 & 0 & 0 & 1 \\
1 & 0 & 0 & 0\\
0 & 0 & 0 & 0
\end{array}\right).
\end{eqnarray}
which yields straightforward expressions for number operators: $d^\dagger_\uparrow d_\uparrow  = \text{diag} (0,1,0,1)$, and  $d^\dagger_\downarrow d_\downarrow  = \text{diag} (0,1,1,0)$. 
The coupling to the normal lead induces the transition between the eigenstates. Since the normal lead is eligible to exchange an electron, to construct the Lindblad operator we need to express $d_{\sigma}^{(\dagger)}$ in terms of eigen operators of the system.  That is, 
$\tilde d_\uparrow =  U_0^\dagger d_\uparrow U_0$, $\tilde d_\downarrow = U_0^\dagger d_\downarrow U_0$. 
The Lindblad operator comes with the coupling strength between the system and the electronic bath,
\begin{equation}
\sqrt{\gamma^e_{N,i\leftarrow j}} =\sqrt{\Gamma_r f_r^{e} (E_i-E_j)},
\end{equation}
where $f^e_{r}(E_i-E_j) = [e^{(E_i-E_j-\mu_r)/T_r}+1]^{-1}$ is the Fermi-Dirac distribution, for a transition of adding an electron to the system that accompanies energy change $(E_i - E_j)$ in the system. For example, the Lindblad operator that removes an spin-up electron from the system to the reservoir is:
\begin{eqnarray}\nonumber
\tilde L_\uparrow =  \left(
\begin{array}{cccc}
0 & 0 & 0 & u_+^*\sqrt{\gamma^h_{+\uparrow}} \\
0 & 0 & 0 & u_-^*\sqrt{\gamma^h_{-\downarrow}} \\
v_+\sqrt{ \gamma^h_{\downarrow +}} & v_-\sqrt{ \gamma^h_{\downarrow -}} & 0 & 0\\
0& 0& 0& 0
\end{array} \right)
\end{eqnarray}
where $\gamma^{e(h)}_{ij}\equiv \gamma_{r,|i\rangle \leftarrow |j\rangle} = \Gamma_r f_r^{e(h)}(E_i - E_j)$ ($e$ for electrons and $h$ for holes) and $i,j \in \{+,-,\downarrow,\uparrow\}$. 
Other Lindblad operators $\tilde L^\dagger_{\uparrow},\tilde L_{\downarrow},\tilde L^\dagger_{\downarrow}$ are expressed similarly. 
The equation of motion of the density matrix is
\begin{equation}\label{qleq}
\frac{d \rho }{d t} = -i [ H_{\mathrm{top}}, \rho ] + \mathcal D [  L_\sigma ] +  \mathcal D [  L^\dagger_\sigma], 
\end{equation}
where the dissipator operator due to the normal lead is
\begin{equation} 
\mathcal D [  L^\dagger_\sigma ] =    L_\sigma \rho    L^\dagger_\sigma - \frac 1 2 \{    L^\dagger_\sigma   L_\sigma , \rho \}.
\end{equation}
Rewriting the equation of motion in the eigen state basis of Hamiltonian,
\begin{align}
U_0^\dagger \frac{d \rho }{d t} U_0  &= U_0^\dagger \left( -i [ H_{\mathrm{top}}, \rho ] + \mathcal D [ L_\sigma ] +  \mathcal D [ L^\dagger_\sigma ] \right) U_0 , \\
 \frac{d \tilde \rho }{d t}  &=  -i [ D, \tilde \rho ] + \tilde{\mathcal D} [ L_\sigma ] +   \tilde{\mathcal D} [ L^\dagger_\sigma ] , 
\end{align}
where 
\begin{align} \nonumber
\tilde {\mathcal D} [ L^\dagger_\sigma ] 
 &= U_0^\dagger {\mathcal D} [ L^\dagger_\sigma ] U_0 =  (U_0^\dagger L_\sigma U_0) \tilde \rho (U_0^\dagger L^\dagger_\sigma U_0) \\  & -
 \frac 1 2 \{  U_0^\dagger (L^\dagger_\sigma L_\sigma ) U_0 , \tilde \rho \}.
 \end{align}
For the derivation of the master equation from the quantum Lindblad equation [Eq. (\ref{qleq})] focusing on the change of the occupation numbers of eigen states and disregarding the coherences. For that issue we employ the density matrix with diagonal elements only $ \tilde \rho = U_0^\dagger \rho U_0  =\text{diag} (\rho_+, \rho_-, \rho_{\downarrow}, \rho_{\uparrow} )$. This approach
is justified as long as $\Gamma_S\ll k_B T_{\alpha}$ ($\alpha\in r,h,c $). Considering this approach we obtain the following set of equation of motions:  
\begin{widetext}
\begin{eqnarray} \label{eq:master1}
&&\frac{\partial \rho_+}{\partial t} =   \left[ |u_+|^2 (\gamma^h_{+,\uparrow}\rho_\uparrow +  \gamma^h_{+,\downarrow}\rho_\downarrow) - |v_+|^2 (\gamma^h_{\downarrow +} + \gamma^h_{\uparrow +}) \rho_+ \right] +   \left[ |v_+|^2 (\gamma^e_{+,\uparrow}\rho_\uparrow + \gamma^e_{+,\downarrow}\rho_\downarrow) - |u_+|^2 (\gamma^e_{\downarrow +} + f^e_{\uparrow +}) \rho_+ \right], 
\\ 
&&\frac{\partial \rho_-}{\partial t} =   \left[ |u_-|^2 (\gamma^h_{-\uparrow}\rho_\uparrow + \gamma^h_{-\downarrow}\rho_\downarrow) - |v_-|^2 (\gamma^h_{\downarrow -} + \gamma^h_{\uparrow -})  \rho_- \right] +   \left[ |v_-|^2 (\gamma^e_{-\uparrow}\rho_\uparrow + \gamma^e_{-\downarrow}\rho_\downarrow) - |u_-|^2 (\gamma^e_{\downarrow -} + \gamma^e_{\uparrow -}) \rho_- \right],
 \\ 
 \nonumber
&&\frac{\partial \rho_\downarrow}{\partial t} =   \left[ |v_+|^2 \gamma^h_{\downarrow +}\rho_+ + |v_-|^2 \gamma^h_{\downarrow -} \rho_- -(|u_+|^2 \gamma^h_{+ \downarrow} + |u_-|^2 \gamma^h_{-\downarrow}) \rho_\downarrow \right] +   [ |u_+|^2 \gamma^e_{\downarrow +}\rho_+ + |u_-|^2 \gamma^e_{\downarrow -} \rho_- - (|v_+|^2 \gamma^e_{+ \downarrow} +|v_-|^2 \gamma^e_{-\downarrow}) \rho_\downarrow ], 
\\
 \\
 \nonumber
&&\frac{\partial \rho_\uparrow}{\partial t} =   \left[ |v_+|^2 \gamma^h_{\uparrow +} \rho_+ + |v_-|^2 \gamma^h_{\uparrow -} \rho_- - (|u_+|^2 \gamma^h_{+ \uparrow} + |u_-|^2 \gamma^h_{-\uparrow}) \rho_\uparrow \right] +   [|u_+|^2 \gamma^e_{\uparrow +} \rho_+ + |u_-|^2 \gamma^e_{\uparrow -} \rho_- 
- (|v_+|^2 \gamma^e_{+ \uparrow} +|v_-|^2 \gamma^e_{-\uparrow}) \rho_\uparrow]. \\ \label{qme_d}
\end{eqnarray}
\end{widetext}
where the amplitude of each term can be read from $\ket \pm = u_\pm \ket{0,0} + v_\pm \ket{1,1} $. When single electron states $\ket {\uparrow}$ or $\ket{\downarrow}$ receives one electron to become $\ket{\pm}$, the transition rate is $|v_\pm|^2$. When single electron states $\ket {\uparrow}$ or $\ket{\downarrow}$ emit one electron to become $\ket{\pm}$, the transition rate is $|u_\pm|^2$. This explains the first two equations with positive signs. The other terms can be read similarly.

\subsection{A double quantum dot}

The extension of the previous derivation of the master equation for the double quantum dot system is straightforward. We simply need to extend the Hilbert space to take into account the occupancy of the bottom quantum dot, $n_b \in \{0,1\}$. Our Hamiltonian is $H_{DQD} = \text{blkdiag}(H_{n_b=0}, H_{n_b=1})$. The first block diagonal Hamiltonian $H_{n_b=0}$ is shown in Eq.~\eqref{Hnb0}. The second block diagonal Hamiltonian is the following: 
\begin{widetext}
\begin{eqnarray} \nonumber
H_\mathrm{n_b=1} = \left(\begin{array}{cccc}
0           &  \Gamma_S                     &  0    & 0 \\
\Gamma_S    &  2\epsilon_t + U_t + 2 U_{tb} &  0    & 0\\
0 & 0       &  \epsilon_t + \epsilon_b + U_{tb} & 0\\
0 & 0       &  0 & \epsilon_t + \epsilon_b + U_{tb} 
\end{array}\right).
\end{eqnarray}
\end{widetext}
with eigenvalues $\epsilon_{\pm}' = \epsilon_t +
U_{tb} + \frac{U_t}{2} \pm \sqrt{\Gamma_S^2 + (\epsilon_t + 
U_{tb} + \frac{U_t}{2})^2}$.
The corresponding eigenvectors are $ \ket{\pm} = u_{\pm} \ket{0,0} + v_{\pm} \ket{1,1}$. By introducing the unitary matrix, 
\begin{eqnarray}
U_{1} = \left(\begin{array}{cccc}
u_{+,1} & u_{-,1} & 0 & 0 \\
v_{+,1} & v_{-,1} & 0 & 0\\
0 & 0 & 1 & 0 \\
0 & 0 & 0 & 1 
\end{array}\right), \,\,\,\,
U_{1}^\dagger =\left(\begin{array}{cccc}
u^*_{+,1} & v^*_{+,1} & 0 & 0 \\
u^*_{-,1} & v^*_{-,1} & 0 & 0\\
0 & 0 & 1 & 0\\
0 & 0 & 0 & 1 
\end{array}\right),\nonumber 
\end{eqnarray}
we have the diagonalization of the Hamiltonian: $
D_{n_b=1} = U_1^\dagger H_{n_b=1} U_1= \text{diag} (\epsilon_+ , \epsilon_-, \epsilon_t + \epsilon_b + U_{tb} , \epsilon_t  + \epsilon_b + U_{tb} )$. 
The Hamiltonian of the double quantum dot is digonalized by the unitary matrix $U = \text{blkdiag} (U_{0}, U_{1})$. 
In the eigen state basis, the Hamiltonian is diagonal: 
$D_{tb} = U^\dagger H_{tb} U = \text{blkdiag}(D_{n_b=0},D_{n_b=1})$. 
We then need to write the equation of motion for the density matrix ignoring the coherence terms
$\rho = \text{diag} (\rho_{+,0}, \rho_{-,0}, \rho_{\downarrow,0},\rho_{\uparrow,0},\rho_{+,1}, \rho_{-,1}, \rho_{\downarrow,1},\rho_{\uparrow,1} )$. 
Because the 8$\times$8 unitary matrix is block diagonal, our previous expressions for the single quantum dot will equally apply for sector $n_b=0$ and $n_b=1$ with modified eigenvalues and eigenvectors. 
Terms newly added are the creation and removal of electrons in the bottom quantum dot. The creation and annihilation operators of an electron in the bottom quantum dot  are expressed as follows
\begin{align}
d^\dagger_b &= \left(\begin{array}{c}
 0 \\
\mathcal I_4  
\end{array}\right), \,\,\,\,\,\, 
d_b =\left( \begin{array}{c}
 \mathcal I_4 \\
 0
\end{array}\right), \\
\tilde d^\dagger_b &= U^\dagger d^\dagger_t U =\left( \begin{array}{cc}
 0&0\\
U_1^\dagger U_0  & 0
\end{array}\right), \\
 \tilde d_b &= U^\dagger  d_b U = \left(\begin{array}{cc}
0 & U_0^\dagger U_1 \\
 0 & 0 
\end{array}\right), 
\end{align}
where the first two are in the electron number basis $\ket{n_\downarrow,n_\uparrow, n_b}$, and the latter two are the expressions in the eigen basis. 
From these, we can build the explicit expression of dissipators from the left and right lead coupled to the bottom quantum dot 

\begin{eqnarray}
\tilde {\mathcal D} [ d^\dagger_b ] \nonumber
& = U^\dagger {\mathcal D} [ d^\dagger_b ] U=  \tilde d_b  \tilde \rho \tilde d^\dagger_b  - \frac 1 2 \{  \tilde d_b^\dagger \tilde d_b, \tilde \rho \}=\\ 
&\left(\begin{array}{cc}
U_0^\dagger U_1 \rho_{n_b=1} U_1^\dagger U_0 & 0\\
0 & -\rho_{n_b=1}
\end{array}\right),
\end{eqnarray}
which corresponds to the physical process of removing one electron from the bottom quantum dot. Thus, it is related to the density matrix sector of $(n_b=1)$. 
The other dissipator is to add a particle to the bottom quantum dot 
\begin{eqnarray}\nonumber
\tilde {\mathcal D} [ d_b ]  & = U^\dagger {\mathcal D} [ d_b ] U =  \tilde d^\dagger_b  \tilde \rho \tilde d_b  - \frac 1 2 \{  \tilde d_b \tilde d^\dagger_b, \tilde \rho \} =\\ 
&\left(\begin{array}{cc}
-\rho_{n_b=0}  &  0\\
0 & U_1^\dagger U_0 \rho_{n_b=0} U_0^\dagger U_1
\end{array}\right),
\end{eqnarray}
From the two dissipators, there are scattering between states $\ket{+,1}$ and $\ket{-, 0}$, for example. Specifically, the diagonal elements are the following: 
\begin{widetext}
\begin{eqnarray}
\left( U_0^\dagger U_1 \rho_{n_b=1} U_1^\dagger U_0 \right)_{11} &= |u_{+,0}^* u_{+,1} + v_{+,0}^* v_{+,1} |^2 \rho_{+,1} + |u_{+,0}^* u_{-,1} + v_{+,0}^* v_{-,1}|^2 \rho_{-,1}, \\
\left( U_0^\dagger U_1 \rho_{n_b=1} U_1^\dagger U_0 \right)_{22} &= |u_{-,0}^* u_{-,1} + v_{-,0}^* v_{-,1} |^2 \rho_{-,1} + |u_{-,0}^* u_{+,1} + v_{-,0}^* v_{+,1}|^2 \rho_{+,1}, \\
\left( U_1^\dagger U_0 \rho_{n_b=0} U_0^\dagger U_1 \right)_{11} &= |u_{+,1}^{*} u_{+,0} + v_{+,1}^{*} v_{+,0} |^2 \rho_{+,0} + |u_{+,1}^{*} u_{-,0} + v_{+,1}^{*} v_{-,0}|^2 \rho_{-,0}, \\
\left( U_1^\dagger U_0 \rho_{n_b=0} U_0^\dagger U_1 \right)_{22} &= |u_{-,1}^{*} u_{-,0} + v_{-,1}^{*} v_{-,0} |^2 \rho_{-,0} + |u_{-,1}^{*} u_{+,0} + v_{-,1}^{*} v_{+,0}|^2 \rho_{+,0}, 
\end{eqnarray}
\end{widetext}
For different scatterings, the associated energy transfer is different. That is, we need to introduce different argument for the Fermi-Dirac distribution that is multiplied for each scattering process. Thus, we need to include the coupling strength between the system and baths $\sim \sqrt{\Gamma_l f^{e,h}_{l,\ket{i} \leftarrow \ket{j}}}$ to complete the expression of the dissipators. As a result, for the double quantum dot system we have additional terms accounting for the scatterings changing the electron number of bottom quantum dot, $n_b$. 
\begin{widetext}
\begin{align} \nonumber
\frac{\partial \rho_{+,0}}{\partial t} = \left.\frac{\partial \rho_{+,0}}{\partial t} \right|_{n_b=0\leftarrow 0}  
& +\sum_{l=h,c} \gamma^h_{l,\ket{+,0} \leftarrow \ket{+,1}} |u_{+,0}^* u_{+,1} + v_{+,0}^* v_{+,1} |^2 \rho_{+,1} + \gamma^h_{l,\ket{+,0} \leftarrow \ket{-,1}} |u_{+,0}^* u_{-,1} + v_{+,0}^* v_{-,1}|^2 \rho_{-,1} 
  \nonumber \\ \nonumber
& - \sum_{l=h,c}  \gamma^e_{l,\ket{+,1} \leftarrow \ket{+,0}}|u_{+,1}^{*} u_{+,0} + v_{+,1}^{*} v_{+,0} |^2 \rho_{+,0} + \gamma^e_{l,\ket{-,1} \leftarrow \ket{+,0}}|u_{-,1}^{*} u_{+,0} + v_{-,1}^{*} v_{+,0}|^2 \rho_{+,0}
, \\ 
\nonumber
\frac{\partial \rho_{-,0}}{\partial t} = \left.\frac{\partial \rho_{-,0}}{\partial t} \right|_{n_b=0\leftarrow 0}
  &+\sum_{l=h,c}    \gamma^h_{l,\ket{-,0} \leftarrow \ket{-,1}}|u_{-,0}^* u_{-,1} + v_{-,0}^* v_{-,1} |^2 \rho_{-,1} + \gamma^h_{l,\ket{-,0} \leftarrow \ket{+,1}})|u_{-,0}^* u_{+,1} + v_{-,0}^* v_{+,1}|^2 \rho_{+,1} \nonumber \\ \nonumber
& -\sum_{l=h,c}  \gamma^e_{l,\ket{-,1} \leftarrow \ket{-,0}}|u_{-,1}^{*} u_{-,0} + v_{-,1}^{*} v_{-,0} |^2 \rho_{-,0}   + \gamma^e_{l,\ket{+,1} \leftarrow \ket{-,0}}|u_{+,1}^{*} u_{-,0} + v_{+,1}^{*} v_{-,0}|^2 \rho_{-,0}, 
\\ \nonumber
 \frac{\partial \rho_{\downarrow,0}}{\partial t} =\left.\frac{\partial \rho_{\downarrow,0}}{\partial t} \right|_{n_b=0\leftarrow 0}   &+\sum_{l=h,c} \gamma^h_{l,\ket{\downarrow,0}\leftarrow \ket{\downarrow,1}} \rho_{\downarrow,1} - \gamma^e_{l,\ket{\downarrow,1} \leftarrow \ket{\downarrow,0}}  \rho_{\downarrow,0},
 \\ 
\frac{\partial \rho_{\uparrow,0}}{\partial t} =  \left.\frac{\partial \rho_{\uparrow,0}}{\partial t} \right|_{n_b=0\leftarrow 0} 
&+ \sum_{l=h,c} \gamma^h_{l,\ket{\uparrow,0} \leftarrow \ket{\uparrow,1}}  \rho_{\uparrow,1} - \gamma^e_{l,\ket{\uparrow,1} \leftarrow \ket{\uparrow,0}}  \rho_{\uparrow,0}, \nonumber \label{qme_dd}
\end{align}
\end{widetext}
where $\left.\frac{\partial \rho_k}{\partial t} \right|_{n_b=0\leftarrow 0} $ are the  same equations as appears in Eq. (\ref{eq:master1}-\ref{qme_d}) with $n_b=0$ in which the transition rates  read as follows: $\gamma^{e(h)}_{ij}=\gamma^{e(h)}_{|i,n_b\rangle \leftarrow |j n_b\rangle}= \Gamma_r f^{e(h)}(E_{i,n_b}-E_{j,n_b})$ with $i,j\in\{\pm,\sm\}$, 
the particle number $n_b$ in the bottom dot does not change. 
We have another set of equation of motion for $n_b=1$: 
\begin{widetext}
\begin{align} \nonumber
\frac{\partial \rho_{+,1}}{\partial t} =  \left.\frac{\partial \rho_{+,1}}{\partial t} \right|_{n_b=1\leftarrow 1}  &+\sum_{l=h,c} \gamma^e_{l,\ket{+,1} \leftarrow \ket{+,0}}|u_+^{'*} u_+ + v_+^{'*} v_+ |^2 \rho_+ + \gamma^e_{l,\ket{+,1} \leftarrow \ket{-,0}}|u_{+,1}^{*} u_{-,0} + v_{+,1}^{*} v_{-,0}|^2 \rho_{-,0} 
\\
  &-\sum_{l=h,c}  \gamma^h_{l,\ket{+,0} \leftarrow \ket{+,1}}|u_{+,0}^* u_{+,1} + v_{+,0}^* v_{+,1} |^2 \rho_{+,1}  +  \gamma^h_{l,\ket{-,0} \leftarrow \ket{+,1}}|u_{-,0}^* u_{+,1} + v_{-,0}^* v_{,1}+|^2 \rho_{+,1}  \nonumber
 , \\
\frac{\partial \rho_{-,1}}{\partial t} = \left.\frac{\partial \rho_{-,1}}{\partial t} \right|_{n_b=1\leftarrow 1}  &+\sum_{l=h,c}   \gamma^e_{l,\ket{-,1} \leftarrow \ket{-,0}}|u_{-,1}^{*} u_{-,0} + v_{-,1}^{*} v_{-,0}|^2 \rho_{-,0} + \gamma^e_{l,\ket{-,1} \leftarrow \ket{+,0}}|u_{-,1}^{*} u_{+,0} + v_{-,1}^{*} v_{+,0}|^2 \rho_{+,0}  \nonumber \\ 
 &-\sum_{l=h,c} \gamma^h_{l,\ket{-,0} \leftarrow \ket{-,1}}|u_{-,0}^* u_{-,1} + v_{-,0}^* v_{-,1} |^2 \rho_{-,1}+   \gamma^h_{l,\ket{+,0} \leftarrow \ket{-,1}}|u_{+,0}^* u_{-,1}+ v_{+,0}^* v_{-,1}|^2 \rho_{-,1},\nonumber
\\
\frac{\partial \rho_{\downarrow,0}}{\partial t} = \left.\frac{\partial \rho_{\downarrow,1}}{\partial t} \right|_{n_b=1\leftarrow 1} &+\sum_{l=h,c} \gamma^e_{l,\ket{\downarrow,1} \leftarrow \ket{\downarrow,0}}  \rho_{\downarrow,0} - \gamma_{l,\ket{\downarrow,0} \leftarrow \ket{\downarrow,1}}^h \rho_{\downarrow,1}, \nonumber
\\
\frac{\partial \rho_{\uparrow,1}}{\partial t} = \left.\frac{\partial \rho_{\uparrow,1}}{\partial t} \right|_{n_b=1\leftarrow 1} &+\sum_{l=h,c} \gamma^e_{l,\ket{\uparrow,1} \leftarrow \ket{\uparrow,0}}  \rho_{\uparrow,0} - \gamma^h_{l,\ket{\uparrow,0} \leftarrow \ket{\uparrow,1}}  \rho_{\uparrow,1}, \nonumber
\end{align}
\end{widetext}
where $\left.\frac{\partial \rho_k}{\partial t} \right|_{n_b=1\leftarrow 1} $ are the  same equations as appears in Eq. (\ref{eq:master1}-\ref{qme_d}) with $n_b=1$.
As a result, we obtain the master equation used in the main text.

\end{document}